\documentstyle[multicol,epsf,aps,prb,amssymb]{revtex}
\begin{document}

\title{Prospects for Non--Fermi--Liquid Behavior of a Two Level Impurity
   in a Metal.}
\author{Aris L. Moustakas and Daniel S. Fisher}
\address{Department of Physics, Harvard University, Cambridge, MA 02138}
\maketitle

\begin{abstract}
 It has been claimed that an impurity hopping between two
  sites in a metal can lead to non--Fermi liquid
 behavior if the bias between the two sites is tuned to zero.
 Here it is shown that several extra tunnelling processes are
 allowed; those involving bare impurity tunnelling as well as
 two--electron--assisted  tunnelling play important roles. These terms
 generally drive the system away from the intermediate coupling
 non--Fermi--liquid fixed point, which has the character of the two
 channel Kondo problem. In order to find the non-Fermi liquid behavior
 a combination of these terms must be small over a range of energy
 scales. The prospects for obtaining these conditions are studied by
 analyzing both the weak coupling and the intermediate coupling
 behavior, together with the connecting crossover
 region. Unfortunately, the conclusions are rather pessimistic. The
 phase diagram and the important crossover are analyzed via
 bosonization and refermionization and some of the results compared to
 those obtained from conformal field theory. Finally, an additional
 non-Fermi liquid fixed point is  found, a multicritical point with
 two relevant directions even in the unbiased case.
\end{abstract}

\pacs{PACS Numbers: 71.27.+a, 72.10.Fk, 73.40.Gk}
%Strongly Correlated Electron Systems; Scattering by point defects,
%dislocations, surfaces and other imperfections (including Kondo
%effect); Tunneling.
\begin{multicols}{2}
\section{Introduction}
\label{Introduction}
A two--channel Kondo impurity in a metal is, perhaps,  the most promising
 impurity system  to exhibit low temperature non--Fermi--liquid behavior
 experimentally. This hope rests on the relatively few  degrees of
 freedom involved (a local spin doublet coupled to two degenerate
 ``flavors'' of spin--$\frac{1}{2}$ fermions) with the symmetry group
 under which they transform being not very complicated
 [$SU(2)_{flavor}\times SU(2)_{spin}\times U(1)_{charge}$]. As a result
 one might hope that it would not be difficult to find a system
 whose low temperature behavior would be described by the two--channel
 Kondo system.

Yet, for more than a decade since it was first
 introduced by Nozi\'{e}res and Blandin\cite{Nozieres1}, no
 experimental realization of this model has been conclusively
 demonstrated. The difficulty lies in the fact that the
 non--Fermi--liquid fixed point is unstable to various
 symmetry--breaking processes which turn out to be present in real
 experimental situations. For example,  Nozi\'{e}res
 and  Blandin\cite{Nozieres1}  pointed out that anisotropy between
 the two flavor channels, caused by lattice effects, would destroy the
 non--Fermi--liquid ground state. Since this anisotropy could not be
 made to vanish for the known cases, the search for
 non--Fermi--liquid behavior with conventional spin--Kondo systems was
 suspended. Cox\cite{Cox} pointed out, however, that under certain
 symmetry conditions, local quadrupolar degrees of freedom could
 result in two channel Kondo-like coupling; unfortunately, dilute
 impurity systems of this type have proven hard to make.

Later, Vlad\'{a}r and Zawadowski\cite{Zawadowski1} suggested that  a
 non--magnetic  impurity tunnelling between two sites in a metal could
 be modeled as a two--channel Kondo system in which the roles of the
 channels and the spins in the original formulation are
 interchanged. In this system the spin of the electron plays the role
 of the ``flavor channels'', so that  the anisotropy between
 ``channels'' is no longer an issue
 since in  zero external magnetic field the spin--up and the
 spin--down electrons are degenerate. This led Vlad\'{a}r and
 Zawadowski\cite{Zawadowski1} to predict
 non--Fermi--liquid behavior in such a system. Recently Ralph
 {\em et al}\cite{Ralph1,Ralph2} have interpreted  low
 temperature tunnelling data in very small metallic contacts in terms
 of two--channel Kondo--like physics. Their  measurements are
 claimed to be consistent with certain exact results obtained by Affleck and
 Ludwig\cite{Affleck1}, but at this point, the interpretation is
 still controversial\cite{Wingreen}.

Unfortunately, the mapping of the two--site impurity to the
 two--channel Kondo (2CK) system is far from exact, and, even with
 no anisotropy between ``channels'' (the spin--up
 and spin--down electrons) there are other processes present in the tunnelling
 impurity system  which are relevant and hence generically destroy the
 non--Fermi--liquid ground state. These processes, which cannot be
 neglected have not been treated adequately in the
 literature\cite{Zawadowski1,Zimanyi1,Murumatsu}.

In this paper we carefully consider the mapping between the tunnelling
impurity system and the two--channel Kondo (2CK) problem. We first
analyze the behavior when the impurity tunnelling starts to become
important as the temperature is lowered; this is  analogous to the
weak coupling regime of the Kondo problem and is needed in order to
understand which tunnelling processes will dominate at low
temperatures. We then analyze the intermediate coupling behavior to
investigate whether there are parameter regimes in which the system
will be governed by the 2CK fixed point over a reasonable range of
temperatures -- as would be needed to observe the non--Fermi liquid
behavior experimentally.

Unfortunately, even in the optimal case of an impurity tunnelling
between two identical sites so that the system has an extra $Z_2$
symmetry, we find that the 2CK fixed point is only accessible if two
tunnelling processes, which are very different physically, nearly
exactly cancel. Generically, the system will exhibit Fermi liquid
behavior at low temperatures.

We analyze the behavior near the 2CK point in detail,
focusing on the connection between the physical operators and those
that appear as ``natural'' operators in the 2CK language. It is found
that, on the critical manifold which can be obtained in the symmetric
impurity problem by adjusting one parameter, there are {\em four}
leading irrelevant directions in contrast to the behavior for the pure
2CK problem analyzed by Sengupta and Georges\cite{Georges}.
Various symmetry breaking terms are also studied and our results
recover those derived by Affleck {\em et al}\cite{Affleck2,Affleck3} with
conformal field theory techniques.

A somewhat surprising feature emerges in the fuller phase diagram of
the symmetric impurity model: a {\em second} fixed point which
exhibits non-Fermi liquid behavior, albeit one with two
relevant directions in the $Z_2$ symmetric case.

\subsection{Outline}
In the remainder of the Introduction we motivate the form of the
Hamiltonian in which we focus and interpret the various terms that
should appear.  In Section II we derive an effective Hamiltonian that
we will study and analyze its symmetries, while in Section III the
weak coupling analysis is outlined. In Section IV the behavior near
the intermediate coupling 2CK fixed point and, for completeness, the
various symmetry breaking operators near the 2CK fixed point are
studied. In Section V we discuss the existence of an extra novel fixed
point at intermediate coupling. In Section VI, we discuss the
accessibility of the 2CK fixed point and draw our conclusions.
Finally, in the Appendices, the details of the weak--coupling
(Appendix A) and the intermediate--coupling analysis (Appendix B) are
presented; the comparison of our results to those obtained by
conformal field theory is made in Appendix C.

\subsection{Physical Picture}
The system we wish to describe is an impurity or heavy particle which
can hop back and forth between two sites coupled to a bath of
electrons. The two sites may or may not be equivalent but we will
primarily focus on the symmetric (equivalent) case. The asymmetric
case can readily be treated in a similar manner. The effects of the
interaction of this impurity with the electrons can be manifested in a
number of ways. First, the electrons will tend to screen the charge of
the impurity. Thus the impurity will hop between the two sites
carrying with it tightly bound electrons which can move fast enough to
adjust to the position of the impurity; it is convenient to consider
these to simply be part of the ``dressed'' impurity particle. However,
in addition, as the impurity moves it may also redistribute the low
energy electronic excitations near the Fermi surface. Since we are
interested in the low energy physics, these processes must be treated
directly. For simplicity we consider only $s$-wave scattering off the
impurity.

If an impurity of charge $Ze$ hops between two well separated sites,
then the Friedel sum rule relates the ($s$-wave) scattering phase
shift off a static impurity, $\delta$, to the electronic charge that
will be moved to screen the impurity as it moves adiabatically from
one site to the other, via $Z=2\delta/\pi$, with the factor of two due
to the two spin species. Conversely, if the two sites are close
together, one can still usefully speak of an effective charge $Q$ (per
spin) which plays an analogous role to $Q=\delta/\pi$ in the well
separated case, but is no longer simply related either to scattering
phase shifts or to the impurity charge. It will instead turn out to be
exactly the ``orthogonality catastrophe'' exponent that determines the
system size dependence of the overlap between the electronic ground
states with the impurity at the two sites.\cite{Yamada} For
simplicity, we will focus on $Q$ in the range $0\leq Q \leq 1$,
corresponding to repulsive interactions. (As shown in reference
\onlinecite{MF}, other ranges of the effective charge can be reduced
to this case via a set of more complicated combined impurity-electron
processes related to those we consider here.)

The important processes, in addition to hopping of  the (dressed)
impurity by itself, will be those in which one or two low energy
electrons move in the {\em opposite} direction to that the impurity
hops. These processes can, for $Q>1/2$,  reduce the effective charge
that must relax to the new impurity position, thereby decreasing the
orthogonality  between the pre- and post-hop configuration; this
results in a larger amplitude for the combined impurity-electron
process at low temperatures, relative to the simple impurity
hop process. [Note that in reference \onlinecite{MF}, a
sign error in the definition of $Q$ resulted in the incorrect
interpretation being given to these processes; this error does not affect
the conclusions, just the interpretation of the combined hopping
processes].

In order to proceed with the analysis of the low temperature behavior
of interest it would appear to be important to assess the relative
magnitudes of the amplitudes, at some
high energy scale, of the processes discussed above. However, it has
been claimed\cite{Kagan}  that this is
essentially impossible for strong electron-impurity
interactions. Indeed Kagan and Prokof'ev\cite{Kagan} have claimed that
a sensible Hamiltonian cannot be written in terms of a simple two
level system since the high energy electronic degrees of freedom
cannot be properly taken into account. Although there are indeed real
difficulties here, it should nevertheless be possible to introduce
{\em effective} amplitudes at some intermediate energy scale and then
analyze the behavior of the system in the phase space of these
effective parameters. In general, unless there are specific reasons to
prevent it, one would expect that most combinations of parameters
could, in principle, be realized. We thus approach the problem via
this route and start with an effective Hamiltonian at an intermediate
energy scale, at, say, some fraction of the conduction bandwidth. As
we shall see, the extra hopping processes will in any case be
generated at low energies.

\section{Effective Hamiltonian}
\label{Effective Hamiltonian}
The important electronic degrees of freedom at low energies are those
that interact with the impurity in one of its two positions ``1'' and
``2''. These are just the $s$-wave conduction electrons around the
positions of the two sites. Thus at each energy, there will be two
important electronic degrees of freedom per spin. However,
unless the two sites are very far apart (in which case the impurity
tunnelling rates will be negligible and hence not of interest), the
two sets of $s$-wave electrons will {\em not} be orthogonal; this will
play an important role in what follows. If we label the two sets of
$s$-wave electrons by their energy, $\epsilon$, measured from the
Fermi surface, then for each $\epsilon$ there is an (essentially)
unique pair of linear combinations of the two $s$-wave states, that
are orthonormal and transform into each other under interchange of the
two sites. We label the annihilation operators of this orthonormal
pair $c_{1\epsilon}$ and
$c_{2\epsilon}$ with anticommutation relations
\begin{equation}
  \label{anticommutationscie}
  \left\{ c_{i\epsilon},c^\dagger_{j\epsilon'}\right\} = 2\pi
    \delta\left(\epsilon-\epsilon'\right) \delta_{ij}
\end{equation}
with the ``1'' and ``2'' denoting the sites near which the
wavefunction is larger. The impurity interacts with, simply, the
operators
\begin{equation}
  \label{defcilocal}
  c_{1,2}= \int \frac{d\epsilon}{2\pi} c_{1,2\epsilon},
\end{equation}
although in each position the impurity will couple to {\em both}
$c_1$ and $c_2$ due to the non-orthogonality of  the two electronic
$s$-wave states. With time reversal invariance, which we assume
henceforth, the most general interaction with the impurity becomes
(ignoring spin for now)
\begin{eqnarray}
  \label{Uinit}
  U= d^\dagger_1d_1 \left[V_1 \left(c^\dagger_1c_1
  +c^\dagger_2c_2\right)\right. &+& V_2
  \left(c^\dagger_1c_2+c^\dagger_2c_1\right) \\ \nonumber
  &+&V_3\left.\left(c^\dagger_1c_1
  -c^\dagger_2c_2\right)\right]  \\ \nonumber
+ d^\dagger_2d_2 \left[V_1 \left(c^\dagger_2c_2
  +c^\dagger_1c_1\right) \right.&+& V_2
  \left(c^\dagger_2c_1+c^\dagger_1c_2\right)  \\ \nonumber
&+&V_3\left.\left(c^\dagger_2c_2
  -c^\dagger_1c_1\right)\right]
\end{eqnarray}
where $d_{1,2}$ are the annihilation operator of the impurity at the sites.
Using the obvious identity $d^\dagger_1d_1+d^\dagger_2d_2=1$, Eq(\ref{Uinit})
can be written as
\begin{eqnarray}
  \label{Ufinal}
   U&=& V_1 \left(c^\dagger_1c_1 + c^\dagger_2c_2\right)
    + V_2 \left(c^\dagger_1c_2 + c^\dagger_2c_1\right) \\ \nonumber
    &+& V_3 \left(d^\dagger_1d_1 -d^\dagger_2d_2\right) \left(c^\dagger_1c_1
  -c^\dagger_2c_2\right).
\end{eqnarray}
Note the appearance of an effective electronic hopping term $V_2$,
caused by the scattering by the impurity; this will vanish if the
sites are far apart, but in general will be comparable to the other
terms.

The first term in Eq(\ref{Ufinal}) which is the average of the
interaction over the two impurity positions, merely produces a constant
phase shift for both ``1'' and``2'' electrons at the Fermi level and,
combined with other operators, gives rise only to irrelevant
terms. Therefore we will ignore it at this point
although in Section IV  an irrelevant operator it
gives rise to will play a role in our discussion of the intermediate
coupling behavior.

With the effective hopping charge $Q$  in the range
$\left[0,1\right]$,  there are three
hopping processes that must be considered\cite{MF}:
\begin{eqnarray}
  \label{Hhop1}
  {\cal H}_{hop} = d^\dagger_2 d_1 \biggl[\Delta_0 +
  \frac{\Delta_1}{2}
  \left(c^\dagger_{1\uparrow}c_{2\uparrow} +
    c^\dagger_{1\downarrow}c_{2\downarrow} \right)\biggr. \\ \nonumber
+\Delta_2\biggl.
c^\dagger_{1\uparrow}c_{2\uparrow}c^\dagger_{1\downarrow}c_{2\downarrow}
\biggr] + h.c.
\end{eqnarray}
representing hopping of the (dressed) impurity, jointly with,
respectively, 0, 1 and 2 electrons moving the opposite way. Although we
might start at a high energy scale with negligible $\Delta_1$ and
$\Delta_2$, these will be generated under renormalization and hence
must be included.

In order to analyze the renormalization group flows, it is convenient
to approximate the conduction band $s$-wave electrons $c_{1\epsilon}$
and $c_{2\epsilon}$ by a linear dispersion with a cutoff, at short
times, $\tau_c$, roughly the inverse bandwidth. Then the interactions
with the impurity will essentially be replaced by corresponding  phase
shifts, specifically $V_3$ replaced by an effective phase shift that
we denote $\pi Q_0$; $Q_0$ will have the interpretation of an
effective ``charge''. We then have, after
inserting powers of $\tau_c$ to make the couplings dimensionless and
factors of $\pi$ for convenience,
\begin{equation}
  \label{calH}
  {\cal H}= {\cal H}_0 + {\cal H}_{int} + {\cal H}_{hop}
\end{equation}
with
\begin{equation}
  \label{calHo}
  {\cal H}_0= \sum_\sigma \int \frac{d\epsilon}{2\pi} \epsilon
  \left[c^\dagger_{1\sigma\epsilon}c_{1\sigma\epsilon} +
   c^\dagger_{2\sigma\epsilon}c_{2\sigma\epsilon}\right],
\end{equation}
\begin{eqnarray}
  \label{calHint}
  {\cal H}_{int} &=& \pi Q_0 \left(d^\dagger_2d_2 -d^\dagger_1d_1\right)
  \sum_\sigma  \left(c^\dagger_{2\sigma}c_{2\sigma}
  -c^\dagger_{1\sigma}c_{1\sigma}\right) \\ \nonumber
 &+& \pi y \sum_\sigma
    \left(c^\dagger_{1\sigma}c_{2\sigma}
  +c^\dagger_{2\sigma}c_{1\sigma}\right)
\end{eqnarray}
and
\begin{eqnarray}
  \label{calHhop}
  {\cal H}_{hop} = d^\dagger_2 d_1 \biggl[\frac{\Delta_0}{2\pi\tau_c}+
   \frac{\Delta_1}{2}
  \sum_\sigma c^\dagger_{1\sigma}c_{2\sigma} \biggr.\\ \nonumber
+\biggl. \Delta_2 2\pi\tau_c
c^\dagger_{1\uparrow}c_{2\uparrow}c^\dagger_{1\downarrow}c_{2\downarrow}
\biggr] + h.c.
\end{eqnarray}
We have rescaled the electronic term $V_2$ to a coefficient $y$ which
will play the role of a ``fugacity'' for electronic hops.

\subsection{Symmetries}
It is important at this stage to examine the symmetries of
Eq(\ref{calH}). In addition to time reversal, conservation of the
electrons $\left(c\rightarrow e^{i\phi} c\right)$, conservation of the
impurity $\left(d\rightarrow e^{i\phi} d\right)$, and $SU(2)$ spin
symmetry, the only other symmetry is interchange of the two sites and
the corresponding electronic states ($1\leftrightarrow 2$). Note,
however, that if the only hopping term had been $\Delta_1$, and if $y$
vanished, there would be an {\em extra} artificial symmetry $d_1
\rightarrow e^{i\phi} d_1$, $c_1
\rightarrow e^{i\phi} c_1$, $d_2\rightarrow d_2$, $c_2 \rightarrow
c_2$ corresponding to conservation of $N_1=d^\dagger_1 d_1 +
  n_{c_1}$ and similarly $N_2$ {\em separately}, where $n_{c_1}$
is the number of the ``one'' electrons which, in the
absence of the channel mixing term $y$, are independent of the ``two''
electrons. As shown in reference \onlinecite{MF}, even if the
electronic states had
been optimally chosen so that there was no mixing of ``one'' and
``two'' electrons at the Fermi energy, the energy dependence of
scattering off the impurity would generate extra mixing terms in
${\cal H}_{int}$, that cannot simply be expressed in terms of $c_1$ and
$c_2$. These would break the artificial symmetry and under
renormalization generate a $y\left(c^\dagger_1 c_2 + h.c.\right)$
mixing term even in the absence of impurity motion. Thus it is best to
include $y$ from the beginning. (The neglected energy dependent
scattering terms will then not play an important role).
In order to understand the difficulties of reaching the 2CK-like fixed
point, this step is {\em crucial}.

The artificial symmetry in the absence of the channel mixing and
$\Delta_0$, $\Delta_2$ terms, corresponds to a conserved pseudo-spin
$N_1-N_2$ which is the sum of the ``$z$-components'' of the impurity
pseudo-spin $d^\dagger_2d_2-d^\dagger_1d_1$ and an electronic
pseudo-spin $n_{c_2} - n_{c_1}$. This pseudo-spin can
play the role of spin for the two-channel Kondo effect and, indeed,
under renormalization the system will flow to this intermediate
coupling 2CK fixed point if $Q>0$.\cite{Footnote1}
Unfortunately, there is no natural small parameter which keeps the
pseudo-spin symmetry breaking terms small.\cite{Footnote2}

\subsection{Bosonization}
In order to carry out the renormalization group analysis for small
bare impurity hopping rates, it is useful, as is standard, to bosonize
the electronic degrees of freedom, treating the electronic states
$c_{1,2 \epsilon}$ as those of a one-dimensional system with two sets
of right moving electrons with ``wavevectors'' $v_F (k-k_F)\propto
\epsilon$. It is simplest to set the Fermi velocity, $v_F=1$, and
treat $\epsilon$ like a
wavevector index, defining
\begin{equation}
  \label{Psij(x)}
 c_{j\sigma}\left(x\right) \equiv \int \frac{d\epsilon}{2\pi}
  e^{i\epsilon x} c_{j\sigma\epsilon}=\frac{1}{\sqrt{2\pi\tau_c}}
  e^{i\Phi_{j\sigma}\left(x\right)}
\end{equation}
so that
\begin{equation}
  \label{dPhidx}
  c^\dagger_{j\sigma}\left(x\right) c_{j\sigma}\left(x\right)=
  \frac{1}{2\pi} \frac{\partial \Phi_{j\sigma}\left(x\right)}{\partial x}
\end{equation}
with $j=1,2$ $\sigma=\uparrow,\downarrow$ and $\Phi_{j\sigma}$ being
the corresponding bosonic degrees of freedom, where we have followed
Emery and Kivelson's notation.\cite{EK} Only $c_{j\sigma}\equiv
c_{j\sigma}\left(x=0\right)$ couples to the impurity. Note that in the
standard expression Eq(\ref{dPhidx}) the left hand side is normal
ordered and therefore the (infinite) uniform charge density does not
appear. Also corrections that vanish as $\tau_c\rightarrow 0$ are
neglected; we will be careful to include the effects of extra terms
when they play an important role.

Since we will later need to be careful to have the proper
anticommutation relations, we must insert extra factors of the form
$\exp\left(i\pi N_\mu\right)$, with $N_\mu\equiv \int dx
\Psi^\dagger_\mu\left(x\right) \Psi_\mu\left(x\right)$, into some of
the bosonized expressions to ensure anticommutations of the different
Fermi fields. These will not play a role as long as no spin-flip
processes occur, and, for the time being we ignore them; the needed
modifications are spelled out in Appendix B.

It is useful to define even and odd components of the Bose fields
\begin{equation}
  \label{defPhieo}
  \Phi_{e,o \sigma}= \frac{1}{\sqrt{2}}\left(\Phi_{2\sigma} \pm
   \Phi_{1\sigma}\right),
\end{equation}
the Hamiltonian then becomes
\begin{eqnarray}
  \label{defbosonizedH}
  {\cal H} &=& {\cal H}_0 + \frac{Q_0}{\sqrt{2}} \sigma_z \sum_\sigma
  \frac{\partial\Phi_{o\sigma}}{\partial x}
  +y\sum_\sigma \cos \Phi_{o\sigma} \\ \nonumber
  &+&\frac{\Delta_0}{2\pi\tau_c} \sigma_x
  + \frac{\Delta_1}{4\pi\tau_c}
  \sum_\sigma \left(\sigma_+
   \exp\left[i\sqrt{2}\Phi_{o\sigma}\right] +
   h.c.\right) \\ \nonumber
   &+&\frac{\Delta_2}{2\pi\tau_c} \left(\sigma_+
   \exp\left[i\sqrt{2}\left(\Phi_{o\uparrow}+\Phi_{o\downarrow}\right)
 \right] + h.c.\right)
\end{eqnarray}
where in all the coupling terms the bosonic fields  are evaluated at
$x=0$ and we  use the impurity
pseudo-spin operators
\begin{equation}
\label{defsigmaz}
\sigma_z=d^\dagger_2d_2-d^\dagger_1d_1
\end{equation}
and
\begin{equation}
\label{defsigma+-}
\sigma_+=d^\dagger_2d_1\; ,\; \sigma_-=d^\dagger_1d_2.
\end{equation}

The conduction electron part of the Hamiltonian can be written in
terms of  boson creation and annihilation operators
$\phi_\mu\left(\epsilon\right)$ with canonical commutation relation
\begin{equation}
  \label{phiphi+comrelation}
  \left[ \phi_\mu\left(\epsilon\right),
    \phi^\dagger_\nu\left(\epsilon'\right) \right] = 2\pi
    \delta_{\mu\nu} \delta\left(\epsilon-\epsilon'\right)
\end{equation}
via
\begin{equation}
  \label{Phimuxasafnofphi}
  \Phi_\mu\left(x\right)=\int_0^\infty
  \frac{d\epsilon}{\sqrt{2\pi\epsilon}} \left[
  \phi_\mu\left(\epsilon\right) e^{i\epsilon
    x}+\phi^\dagger_\mu\left(\epsilon\right) e^{-i\epsilon x}\right]
  e^{-\frac{\epsilon\tau_c}{2}}
\end{equation}
and
\begin{equation}
  \label{Hoasafnofphi}
  {\cal H}_0 = \sum_\mu \int_o^\infty \frac{d\epsilon}{2\pi} \epsilon
\phi^\dagger_\mu\left(\epsilon\right)\phi_\mu\left(\epsilon\right)
e^{-\epsilon\tau_c}
\end{equation}
which involves positive energy parts only. Here  $\tau_c^{-1}$ is the
energy cutoff and $\mu$ represents the various Bose fields, i.e. for
Eq(\ref{defbosonizedH}), $\mu=\left(e\uparrow, e\downarrow, o\uparrow,
o\downarrow\right)$. We see that Eq(\ref{defbosonizedH}) does not
include any terms  with
$\Phi_{e\downarrow}$ or $\Phi_{e\uparrow}$. Thus, up to operators that
are irrelevant for weak coupling, the impurity is decoupled from the
even boson  fields.

It is convenient, following Emery and Kivelson\cite{EK}, to decompose
the odd field $\Phi_o$ which couples to the impurity, into a spin and
charge part, by
\begin{equation}
  \label{Phiosigma}
  \Phi_{o\sigma}=\frac{1}{\sqrt{2}} \left(\Phi_c + \sigma \Phi_s\right)
\end{equation}
with $\sigma=\pm$ for spin $\uparrow$, $\downarrow$, respectively. The
second term in Eq(\ref{defbosonizedH}) becomes simply
$Q_0 \sigma_z \frac{\partial\Phi_c}{\partial x}\left(0\right)$. This
term, which represents the difference in phase shifts for electron
scattering off the two positions of the impurity, can be shifted away
by a unitary transformation which changes the naive weak coupling scaling of
the hopping  terms; this is the conventional approach
used\cite{Murumatsu,Georges,MF,EK} to derive the  weak
coupling flows discussed in the next section.

Although the even fields will not play much role for the time being,
for later purposes we also introduce even fields $\Phi_{ec}$,
$\Phi_{es}$ by
\begin{equation}
  \label{Phiec,s}
  \Phi_{e\sigma}=\frac{1}{\sqrt{2}} \left(\Phi_{ec} + \sigma \Phi_{es}
\right).
\end{equation}
Note that any sum or difference of any two of the $\Phi_{j\sigma}$,
i.e. those that appear from operators bilinear in electron operators,
can be written as a sum or difference of two of the fields $\Phi_s$,
$\Phi_c$, $\Phi_{es}$, $\Phi_{ec}$ with coefficients of {\em unity};
this enables the method of Emery and Kivelson\cite{EK} to work.

\section{Weak hopping analysis}
\label{sec:Analysisoftheweakcouplingpoint}
In order to connect the various amplitudes at the  relatively high energy
scale of  the effective Hamiltonian
(Eq(\ref{defbosonizedH})) to their renormalized values at low energies
 we must
analyze the weak coupling renormalization group (RG) flow
equations for the amplitudes in Eq(\ref{defbosonizedH}). The
magnitudes of the various terms at the crossover scale to intermediate
coupling will determine which regions of the initial parameter space
can flow near to the 2CK fixed point.

Following the procedure in reference \onlinecite{MF} we transform ${\cal H}$
to $U{\cal H}U^\dagger$  using the unitary operator
\begin{equation}
  \label{UtransQo}
  U=\exp\left[-i\sigma_z Q_0 \Phi_c\right]
\end{equation}
Subsequently we follow the RG approach described there and obtain
the following flow equations for the various amplitudes, where for
later convenience we introduce
\begin{equation}
  \label{defq}
q=\frac{1}{2} -Q_0,
\end{equation}
which lies in the range $(-1/2,1/2)$:
\begin{eqnarray}
  \label{weakrgeqns}
  \frac{d\Delta_0}{dl}&=&\left(\frac{1}{2}+2q-2q^2\right)\Delta_0 +
  y\Delta_1 +O\left(\Delta^3\right) \\ \nonumber
  \frac{d\Delta_2}{dl}&=&\left(\frac{1}{2}-2q-2q^2\right)\Delta_2 +
  y\Delta_1 +O\left(\Delta^3\right) \\ \nonumber
  \frac{d\Delta_1}{dl}&=&\left(\frac{1}{2} -2q^2\right)\Delta_1 +
  2y\left(\Delta_0+\Delta_2\right) +O\left(\Delta^3\right) \\
  \nonumber
\frac{dq}{dl}&=&-2q\left(\Delta_0^2 +\Delta_2^2-\frac{1}{2} \Delta_1^2
\right)+ \left(\Delta_0^2 -\Delta_2^2\right) \\ \nonumber
\frac{dy}{dl}&=&\Delta_1\left(\Delta_0+\Delta_2\right)
\end{eqnarray}
The important cross-terms in the first three equations in
Eq(\ref{weakrgeqns}) that are proportional to the electronic mixing
term $y$, have a simple physical interpretation: they represent the
effects of an impurity and an electronic hop both occuring within a
short time interval so that, at lower energies, this appears as simply
the corresponding combined process. Note that we have not included
$O\left(y^2\right)$ terms in the above equations; the definition of
these will depend on the RG procedure, and they will not qualitatively
change the behavior. Thus, in the spirit of focusing on the important
processes and terms, we ignore them\cite{Footnote12A}.

Noting that $q$ and $y$ are constant to order
$O\left(\Delta^2\right)$, to analyze the flow for weak hopping, we can
safely set them to their initial values, $q_0$ and $y_0$. Then the
first three equations can be diagonalized exactly; the details are
discussed in Appendix A. The RG eigenvalues for the
hopping terms, about the zero hopping fixed line are
\begin{eqnarray}
  \label{rgivalues}
  \lambda_\pm&=&\frac{1}{2} -2q^2_0 \pm 2\sqrt{q_0^2+y_0^2} \\ \nonumber
  \lambda_0&=&\frac{1}{2} -2q^2_0.
\end{eqnarray}
We now note that for $0\leq Q_0\leq 1$, corresponding to
$\left|q_0\right|\leq \frac{1}{2}$, at least two eigenvalues are
positive so that impurity hopping is always relevant (leading to the
conclusion of {\em absence} of impurity localization\cite{MF});
likewise for other ranges of $Q_0$, there will always be at least two
relevant hopping processes if there is only $s$-wave scattering off
the impurity.

The Kondo temperature, $T_K$,  is the energy scale at which the first
of the
impurity hopping processes becomes of order unity-- i.e. of order the
renormalized bandwidth.

The system considered by Vlad\'{a}r and
Zawadowski\cite{Zawadowski1} and Vlad\'{a}r {\em et
  al}\cite{Zimanyi1}, essentially
amounts to neglecting $\Delta_2$ and $y$; for small
$\left|q\right|$, which will turn out to be the most interesting case,
this misses part of the physics. The reason is simply that they
neglect one relevant operator $\Delta_2$ which mixes with the other
two to give the correct eigenvalues
(Eq\ref{rgivalues}). Furthermore, as will be seen later,
non-vanishing values of $\Delta_2$ and $y$ are crucial to give the
correct renormalization flows close to the intermediate coupling fixed
point.

If we only kept $Q$ and $\Delta_1$  non-zero, their weak coupling flows
would be   (up to
coefficients) like those for $J_z$ and $J_\perp$ for the conventional
Kondo problem. The Kondo scale is then simply
\begin{equation}
  \label{tkondo}
  \frac{T_K}{W} \sim  \left(\frac{\Delta_1}{W}\right)^\frac{1}{\lambda_0},
\end{equation}
after reinserting factors of the bandwidth, $W$.
(For the special value $\Delta_1=2Q$, $T_K\sim e^{-\frac{1}{\Delta_1}}$
like in the well known anti-ferromagnetic Heisenberg Kondo
problem). For this artificial case, at scales below $T_K$ the novel two
channel Kondo physics will indeed appear, as we will discuss later. This
results from the approach, at low energies, of the system to the
intermediate coupling 2CK fixed point.

Unfortunately, the breaking of the artificial pseudo-spin symmetry
leads to the appearance of terms which generically drive the flow {\em
  away} from the 2CK fixed point. In order to analyze whether the
system can get near to  the 2CK fixed point --- the prerequisite for
observation of non-Fermi liquid behavior---, we must be able to identify
the operators near the 2CK fixed point in terms of the original terms
in the
Hamiltonian; the magnitude of the operators, in particular the relevant
ones,  can then be
determined, roughly, by ``matching'' the coefficients at the
crossover scale,
$T_K$, between the weak and intermediate coupling regimes.

{}From Appendix A, we see that the crossover temperature, $T_K$, will
generally be a complicated function of the original parameters. Before
examining the magnitude of the various important terms at $T_K$, we
turn to the
behavior near the 2CK fixed point; this will tell us which terms need
to be small at scales of order $T_K$ for 2CK behavior to obtain.

\section{Intermediate coupling analysis and two channel Kondo fixed point}
\label{sec:Intermediatecouplingfixedpoint}
{}From the weak coupling flow equations in Eq(\ref{weakrgeqns}), it is
apparent that, in the absence of electronic mixing ($y$=0) there is a
special value of the effective impurity charge $Q$: for
$Q=\frac{1}{2}$, corresponding to $q=0$, $\Delta_0$, $\Delta_1$ and
$\Delta_2$ all scale in the same way with eigenvalue
$\lambda=\lambda_0=\lambda_\pm=\frac{1}{2}$. By analogy to the
Toulouse limit of the conventional one channel Kondo problem, it is
thus natural to look for a solvable point that corresponds to $y=q=0$,
inspired by the observation that free Fermi fields have scaling
dimension of $\frac{1}{2}$ and thus might be used to represent all of
the hopping terms that appear for $q=0$. This has been carried out
recently by Emery and Kivelson\cite{EK} who ``refermionize'' the
bosonized operators that appear in the Hamiltonian enabling the
computation of the scaling of the various important operators.

By examining the weak coupling flows, it is apparent that
special behavior might occur when $\Delta_0+\Delta_2=0$.
Defining
\begin{equation}
  \label{defDelta+}
 \Delta_+\equiv \Delta_0+\Delta_2,
\end{equation}
 we see that, at least to the order included in Eq(\ref{weakrgeqns}),
 for $y=0$ and $\Delta_+=0$, $q$ flows to zero, $y$ is {\em not}
 generated and one might hope that the flow would be towards the 2CK
 fixed point. Indeed, the intermediate  coupling analysis shows that
 this can occur, even if
 \begin{equation}
   \label{defDelta-}
  \Delta_- \equiv
\Delta_0-\Delta_2
 \end{equation}
is non-zero so that the artificial pseudo-spin symmetry is broken.

Physically the role of $\Delta_+$ is very surprising. The processes
represented by $\Delta_0$ and $\Delta_2$ are very different and the
definitions which make them dimensionless are clearly cutoff
dependent. Thus the special critical behavior must not in general
occur exactly
at $\Delta_+=0$, since  the location --- but not the existence --- of
the critical manifold will be affected by irrelevant operators. In
particular, from the weak coupling flows we can see that a non-zero
$q$ combined with $\Delta_- \neq 0$ {\em will} generate $\Delta_+$,
thus a ``bare'' $\Delta_+$ that is non-zero will be needed for the
flow to go to the 2CK fixed point asymptotically.

In order to understand the intermediate coupling behavior and the
special role of $Q=\frac{1}{2}$, following Emery and Kivelson\cite{EK}
and the analogous Toulouse limit\cite{Toulouse} of the conventional
Kondo problem, we perform a unitary transformation with
\begin{equation}
  \label{Utrans1/2}
  U=\exp\left[-\frac{i}{2}\sigma_z\Phi_c\right]
\end{equation}
which transforms the Hamiltonian to $\tilde{\cal H}=U{\cal H}U^+$
\begin{eqnarray}
  \label{Hbosrotated}
   \tilde{\cal H} = {\cal H}_0 &+&
     \frac{\Delta_1}{2\pi\tau_c}\sigma_x\cos\Phi_s +
\frac{\Delta_-}{2\pi\tau_c}\sigma_y\sin\Phi_c  \\ \nonumber
  &+&\frac{\Delta_+}{2\pi\tau_c}\sigma_x\cos\Phi_c
  + 2y\cos\Phi_c\cos\Phi_s -
q\sigma_z \frac{\partial\Phi_c}{\partial x} \\ \nonumber
&+&\frac{u}{\pi}\frac{\partial\Phi_{ec}}{\partial x}
-\frac{w}{2\pi}\sigma_x\cos\Phi_c\frac{\partial\Phi_{ec}}{\partial x}
\end{eqnarray}
where we have combined the terms
\begin{eqnarray}
  \label{explanatoryforHbosrotated}
  \frac{\Delta_2}{2\pi\tau_c}\left(\sigma_+ e^{i\Phi_c} + \sigma_-
  e^{-i\Phi_c}\right) &+&
\frac{\Delta_0}{2\pi\tau_c}\left(\sigma_+ e^{-i\Phi_c} + \sigma_-
  e^{i\Phi_c}\right) \nonumber \\
= \frac{\Delta_-}{2\pi\tau_c}\sigma_y\sin\Phi_c &+&
\frac{\Delta_+}{2\pi\tau_c}\sigma_x\cos\Phi_c,
\end{eqnarray}
and abbreviated $\Phi_\mu\left(x=0\right)$ simply by
$\Phi_\mu$. We have also reintroduced some of the coupling terms to
the even fields, in particular the marginal term
$u\frac{\partial\Phi_{ec}}{\partial x}$ that arises from the
impurity--position independent part of the electron--impurity
scattering ($V_1$ in Eq(\ref{Ufinal})) and the term
$w\sigma_x\cos\Phi_c\frac{\partial\Phi_{ec}}{\partial x}$, which arises
from the combination of $u$ and impurity hopping terms $\Delta_+$ and
is irrelevant for weak hopping, will play roles in our analysis. An
additional irrelevant term, $\sigma_y \sin\Phi_s
\frac{\partial\Phi_{es}}{\partial x}$ couples the impurity to the electronic
{\em spin} degrees of freedom, but does not feed back to the other
operators and thus we ignore it for the time being. Its effect will be
discussed further in Appendix C.

\subsection{Symmetries}
Since  we expect  that the symmetries will play an important role, we
should examine what the original symmetries correspond to in
$\tilde{\cal H}$ and ensure that there are no extra symmetries which
might have arisen from the discarding of operators which were naively
irrelevant, since, as shown in reference \onlinecite{MF},
such procedures,  especially
when combined with ``large''  transformations, such as
Eq(\ref{Utrans1/2}), can be dangerous.

\subsubsection{Gauge invariance and spin conservation}
Since there are no spin-flip processes, separate gauge transformations
can be made for each spin
$\Phi_{j\sigma}\left(x\right)\rightarrow\Phi_{j\sigma}\left(x\right) +
\theta_\sigma$, corresponding to $\Phi_{e\sigma}\rightarrow\Phi_{e\sigma} +
\sqrt{2} \theta_\sigma$. This does not play much role as the
$\Phi_{ec}$-field enters only as a derivative
$\frac{\partial\Phi_{ec}}{\partial x}$ and  the
$\Phi_{es}$-field decouples from the impurity at the level at
which we work (there is feedback, under renormalization from other
operators involving $\Phi_e$ but these only modify pre-existing terms
in $\tilde{\cal H}$).
However, if the $z$-component of electron spin is {\em not} conserved,
then only the  symmetry
$\Phi_{ec}\rightarrow\Phi_{ec} + \theta_{ec}$ remains.

\subsubsection{Periodicity}
The definition of the Fermi fields, Eq(\ref{Psij(x)}), implies that
shifting any $\Phi_{j\sigma}$ by $2\pi$ should leave $\tilde{\cal H}$
unchanged. Depending on whether one or both spin components are
shifted, this implies that (ignoring shifts in $\Phi_{ec}$)
\begin{equation}
  \label{Phicchange}
  \Phi_c\rightarrow\Phi_c +2\pi
\end{equation}
and
\begin{equation}
  \label{Phischange}
  \Phi_s\rightarrow\Phi_s +2\pi
\end{equation}
are independent symmetries, as is the combination
\begin{eqnarray}
  \label{combination1}
  \Phi_c\rightarrow\Phi_c +\pi \;&,&\; \Phi_s\rightarrow\Phi_s +\pi\\
  \nonumber
\sigma_x\rightarrow-\sigma_x \;&,&\; \sigma_y\rightarrow-\sigma_y ;
\end{eqnarray}
with the necessity for the simultaneous transformation of
$\sigma_{x,y}$ resulting from the unitary transformation, U, which
involves $\Phi_c$ so that $\Phi_c\rightarrow\Phi_c +\pi$ introduces an
extra \mbox{$\exp\left(-\frac{\pi i}{2}\sigma_z\right)=-i\sigma_z$} factor
into $U$ yielding $\sigma_{x,y}\rightarrow -\sigma_{x,y}$ in
$\tilde{\cal H}$.

\subsubsection{Interchange}
Interchanging sites one and two is equivalent to
\begin{eqnarray}
  \label{combination2}
  \Phi_c\rightarrow-\Phi_c \;&,&\; \Phi_s\rightarrow-\Phi_s, \\
  \nonumber
  \sigma_y\rightarrow-\sigma_y \;&,&\; \sigma_z\rightarrow-\sigma_z
\end{eqnarray}
with $\Phi_{ec,es}$ unchanged.

\subsubsection{Spin reversal}
Flipping electron spins is simply $\Phi_s\rightarrow-\Phi_s$ and
$\Phi_{es}\rightarrow -\Phi_{es}$.

\subsubsection{Time reversal}
Time reversal transformations change ingoing to outgoing waves,
thereby yielding
$x\rightarrow-x$, $i\rightarrow -i$, all \mbox{$\Phi_\mu\left(x\right)
\rightarrow-\Phi_\mu\left(-x\right)$} and
$\sigma_y\rightarrow-\sigma_y$. Note that here we are {\em not} time
reversing the spins.

\subsubsection{Artificial extra symmetries}
We now see that, indeed, $\tilde{\cal H}$ in Eq(\ref{Hbosrotated})
with all coefficients non-zero, does {\em not} have any artificial
symmetries.
But as seen earlier, an artificial extra pseudo-spin symmetry is possible:
pseudo-spin conservation mod 2 corresponds to $\Phi_c\rightarrow\Phi_c
+\pi$, and more generally full pseudo-spin symmetry corresponds to
independence of the ``one'' and ``two'' electrons,
i.e. \mbox{$\Phi_c\rightarrow \Phi_c +\theta_c$} with any
$\theta_c$. The terms $w$,
$\Delta_+$, $\Delta_-$ and $y$ all violate this, and it can readily be
seen in the representation of $\tilde{\cal H}$  of
Eq(\ref{Hbosrotated}) that $y$ combined with
$\Delta_1$ generates $\Delta_+$, and $q$ combined with $\Delta_-$
generates $\Delta_+$, as expected.

In the representation of Eq(\ref{Hbosrotated}) we see that there is
another possible artificial symmetry: If $w$, $\Delta_+$, $y$ and $q$ are
all zero, then
\begin{equation}
  \label{Phictrans}
  \Phi_c\rightarrow\pi-\Phi_c
\end{equation}
becomes a symmetry. This, as we shall see, restricts the system
automatically to the stable critical manifold of the 2CK fixed point.
But note that because of the unitary transformation of
Eq(\ref{Utrans1/2}) $\Phi_c\rightarrow\pi-\Phi_c$ does {\em not}
correspond to a realizable symmetry in terms of the original variables
since it mixes hops involving the impurity alone and those involving
the impurity together with two electrons. (Indeed many other
irrelevant terms neglected in $\tilde{\cal H}$ will also violate this
artificial symmetry.)

\subsection{Refermionization}
If $\Delta_+$, $q$, $y$ and the other operators neglected in
$\tilde{\cal H}$ all vanish, then the system will still exhibit the
novel intermediate coupling 2CK behavior. As we shall see, the only
relevant operator near the 2CK fixed point, which is consistent with
the true symmetries of the impurity hopping between two equivalent
sites, is $\Delta_+$, thus only one combination of physical quantities
needs to be adjusted to obtain the 2CK behavior. Unfortunately, this
is a combination which is not naturally small.

The behavior near the 2CK fixed point can most easily be found
following Emery and Kivelson\cite{EK} by ``refermionizing'' the Bose
fields $\Phi_c$, $\Phi_s$ and $\Phi_{ec}$ that appear in $\tilde{\cal
  H}$ in Eq(\ref{Hbosrotated}), noting that $\exp i\Phi_\mu$, (with
$\Phi_\mu$ properly normalized) is like some pseudo-Fermi field. The
details are discussed in Appendix B.

Crudely, for $\mu=c$, $s$, $ec$, $es$, each field $e^{i\Phi_\mu}$ is
replaced by a new Fermi field, $\Psi_\mu$, and $\sigma_-$ by a local
Fermi field $d$, with appropriate factors of $e^{i\pi N_\mu}$ and
$e^{i\pi N_d}$ to give the correct anticommutation relations. The
symmetries can most easily be seen, and the Hamiltonian simplified, by
writing the new Fermi fields in terms of a set of Majorana (hermitian)
fermions:
\begin{eqnarray}
  \label{defMajoranafermions}
  d&=&\frac{1}{\sqrt{2}} \left(\gamma+i\delta\right) \\ \nonumber
  \Psi_\mu\left(x=0\right)&=&\frac{1}{\sqrt{2}}\left(\alpha_\mu+
  i\beta_\mu\right).
\end{eqnarray}
Note that
\begin{equation}
  \label{defsigmazwrtgammadelta}
  \sigma_z=2i\gamma\delta.
\end{equation}
The symmetry restrictions can now be examined in terms of these
variables; the details are given in Appendix B.  The periodicity of
the Bose fields $\Phi_c$ and $\Phi_s$ simply implies that only terms
with an even number of Fermi fields can appear in the Hamiltonian.
{\em Gauge invariance} implies that $\Psi_{ec}$ can only appear as
$\Psi_{ec}^\dagger\Psi_{ec}$ and the {\em $z$-component of spin
  conservation} that $\Psi_{es}$ cannot appear in the absence of
magnetic fields in the $x$- or $y$-direction. {\em Spin reversal},
because of the role of the ordering operators, takes
$\Psi_s\rightarrow-\Psi_s^\dagger$ and
$\Psi_{es}\rightarrow-\Psi_{es}^\dagger$, implying that $\alpha_s$ and
$\alpha_{es}$ by themselves are excluded by spin reversal symmetry.
{\em Interchange symmetry} takes $\Psi_c\rightarrow - \Psi_c^\dagger$,
$\Psi_s\rightarrow -\Psi_s^\dagger$ and $d\rightarrow d^\dagger$
thereby requiring that $\alpha_c$ and $\delta$ must appear together.
Finally, {\em time reversal} takes $x\rightarrow -x$, $i\rightarrow
-i$ and $\Phi\rightarrow -\Phi$, allowing only real coefficients of
$\Psi_\mu$ operators, and hence forbidding terms like $i\gamma\alpha$.
The Hamiltonian becomes
\begin{eqnarray}
  \label{Hrefermionized1}
  \hat{\cal H}= {\cal H}_0 &+& \frac{i}{\sqrt{2\pi\tau_c}}
  \left(\Delta_1 \gamma \beta_s + \Delta_-
  \gamma \beta_c + \Delta_+ \delta\alpha_c\right) \\ \nonumber
 &-& 4\pi y \gamma\delta\alpha_c\beta_s + 4\pi q
 \gamma\delta\alpha_c\beta_c\\ \nonumber
 &+&  2iu\alpha_{ec}\beta_{ec} +
w\sqrt{2\pi\tau_c}\delta\alpha_c\alpha_{ec}\beta_{ec}
\end{eqnarray}
with ${\cal H}_0$ the kinetic energy of the four (eight Majorana) new
Fermi fields, $\Psi_\mu$. With the full symmetries of the system, the
other five fields ($\alpha_s$ and the $ec$-, $es$-fields), cannot
appear in the couplings to the
impurity, except in relatively innocuous forms, involving the simple
potential coupling to the average position of the impurity,
$i\alpha_{ec}\beta_{ec}$ and combinations of this with other terms, as
well as the irrelevant term $\delta\alpha_s\alpha_{es}\beta_{es}$,
which will be discussed in Appendix C.
Note that other potentially important operators, like
$\delta\beta_s\alpha_c\beta_c$ and $i\delta
\frac{\partial\alpha_c\left(0\right)}{\partial x}$ are excluded by
time reversal invariance.

The original 2CK problem studied by Emery and Kivelson\cite{EK} and
 Sengupta and Georges\cite{Georges} corresponds to
$\Delta_-=\Delta_+=y=w=0$. In our case non-zero $\Delta_-$ can be
important by observing that ``half'' of the impurity, $\gamma$,
couples to both $\beta_c$ and $\beta_s$, thus it is convenient to
rediagonalize and make linear combinations of these, $\beta_I$ and
$\beta_X$ yielding,  with
\begin{equation}
  \label{defDeltaK}
  \Delta_K=\sqrt{\Delta_1^2 + \Delta_-^2},
\end{equation}
\begin{eqnarray}
  \label{Hrefermionized2}
  \hat{\cal H}= {\cal H}_0 &+&
  \frac{i}{\sqrt{2\pi\tau_c}} \left(\Delta_K
   \gamma \beta_I +
  \Delta_+ \delta\alpha_c \right)\\ \nonumber
 &+& 4\pi\bar{q} \gamma\delta\alpha_c\beta_X +4\pi\bar{y}
 \gamma\delta\alpha_c\beta_I \\ \nonumber
  &+& 2iu\alpha_{ec}\beta_{ec}  +\sqrt{2\pi\tau_c}w
 \delta\alpha_c\alpha_{ec}\beta_{ec},
\end{eqnarray}
where $\bar{y}$ and $\bar{q}$ are linear combinations of the original
$y$ and $q$ (see Eq(\ref{defqybar}) in Appendix B for details). The
above rediagonalization  of $\beta_c$ and $\beta_s$ roughly
corresponds to a rotation of ``spin'' axes in the conventional 2CK
language.

{}From the electronic kinetic energy ${\cal H}_0$, the $\alpha$'s and
$\beta$'s all scale, with time scale $\tau$, as
$\tau^{-\frac{1}{2}}$. If all the couplings are
small, then the anti-commutation relations of $\gamma$ and $\delta$
imply that they are dimensionless so that $\Delta_+$ and $\Delta_K$
scale as $\tau^{-\frac{1}{2}}$, while $\bar{q}$ and $\bar{y}$  are
marginal as from the weak coupling
analysis of Section II (Eq(\ref{weakrgeqns})) and $w$ is irrelevant.

\subsection{Two channel Kondo fixed point and flows}
When $\bar{y}=\bar{q}=\Delta_+=w=0$,
the Hamiltonian in Eq(\ref{Hrefermionized2}) corresponds to the Toulouse
limit analyzed by Emery and Kivelson\cite{EK}. As a free fermion
system it can be analyzed straightforwardly. In this limit ``half'' of
the impurity, $\delta$, is uncoupled from the electrons and thus has
no dynamics, while the other ``half'', $\gamma$, gets dynamics from
coupling to the electrons with  correlations at large imaginary times
\begin{equation}
  \label{gammacorrelation}
  \left<T_\tau\gamma\left(\tau\right)\gamma\left(0\right)\right> \sim
    \frac{1}{\tau}.
\end{equation}
Together these yield the non-Fermi liquid 2CK behavior
\begin{equation}
  \label{sigmazcorrelation}
   \left<T_\tau\sigma_z\left(\tau\right)\sigma_z\left(0\right)\right> \sim
    \frac{1}{\tau}.
\end{equation}
It is important to note that\cite{NozieresPrivComm} the solvable
Hamiltonian is {\em not} generally at the 2CK fixed point. Indeed, the
correlations are readily seen to exhibit crossover from weak coupling
behavior, $\left<\sigma_z \sigma_z\right>\sim const$, to the non-Fermi
liquid behavior of Eq(\ref{sigmazcorrelation}) for
$\tau\gtrsim\Delta_K^{-2} \tau_c$.

The 2CK fixed point, formally, corresponds to
$\Delta_K\rightarrow\infty$. It is more convenient, however, to allow
instead the normalization of $\gamma$ to change, corresponding to
letting the coefficient of the $\int \gamma\partial_\tau \gamma \:
d\tau$ in the Lagrangian vary. At the fixed point, this coefficient,
say $g_\gamma$, will be zero, while $\Delta_K$ becomes a constant; the
correlations of $\gamma$ are then simply the inverse (in frequency
space) of those of $\beta_I$, i.e. a pure power law Eq(\ref{Ggamma}).
To connect the two regimes together, one could choose to renormalize
so that, for example, $\frac{\Delta_K^2}{4\pi}+ g_\gamma=1$ [a
particularly convenient choice; (see Eq(\ref{DeltaKgconstraint})], by
rescaling $\gamma$ under renormalization by a $\Delta_K$-dependent
amount. Details about this procedure are given in Appendix B. [Note,
however, that the resulting fixed point Hamiltonian (with
$\Delta_K=\sqrt{4\pi}$) will {\em not } have the pseudo-spin $SU(2)$
symmetry of the pure 2CK problem. This is because the RG approach we
have implemented here, including the unitary transformation of
Eq(\ref{Utrans1/2}), is inherently anisotropic in channels (equivalent
to spin of conventional 2CK problem). In Appendix C we will show that
this anisotropy will not affect the results.]

At the 2CK fixed point, the above scaling implies
$\gamma\sim\tau^{-\frac{1}{2}}$, while $\delta$ is still dimensionless
so that the RG eigenvalues of the other operators can be read off
immediately: +1/2 for $\Delta_+$, the unique relevant operator
consistent with the symmetries of the problem; -1/2 for $\bar{q}$,
$\bar{y}$ and $w$, which are the three leading irrelevant operators
discussed so far (the fourth will appear in Appendix C); and 0 for
$u$, which is marginal but redundant, in that it does not affect the
impurity dynamics.

The operator corresponding to $\bar{q}$ does not give rise to any
terms which couple $\delta$ linearly to the $\alpha$'s and $\beta$'s
and hence will not generate $\Delta_+$. It
is the leading irrelevant operator for the 2CK problem identified by
 Sengupta and Georges\cite{Georges}. An artificial symmetry is
responsible for its special role, indeed just  the one discussed
earlier: Eq(\ref{Phictrans}),
$\Phi_c\rightarrow \pi-\Phi_c$ corresponds to the discrete symmetry
$\alpha_c\rightarrow-\alpha_c$ and $\beta_X\rightarrow-\beta_X$ which
does not have a natural representation  even in terms of $\Phi_c$ and
$\Phi_s$. But if present, this artificial symmetry  excludes the
generation of terms like $\Delta_+$, $\bar{y}$ and $w$,  even in the
presence of $\bar{q}$. In fact, without these terms, the extra
artificial symmetry is really an  $O(2)$ symmetry in the
$\{\alpha_c, \beta_X\}$ pair, consisting of a  $U(1)$ of
rotations in the $\left(\alpha_c, \beta_X\right)$ plane, combined (in a
non-commutative way) with $Z_2$, the usual site interchange symmetry, which
takes $\alpha_c\rightarrow-\alpha_c$ and
$\beta_X\rightarrow\beta_X$. This is exactly analogous to the $O(2)$
symmetry present in the  model treated by
Emery and Kivelson\cite{EK} and
Sengupta and Georges\cite{Georges}.

In contrast, the irrelevant operators $w$ and $\bar{y}$, break the
artificial symmetry and yield, at lowest order, the generation of the
relevant operator $\Delta_+$ (see Eq(\ref{intermediatergeqns}))
\begin{equation}
  \label{Delta+rgeq}
  \frac{d\Delta_+}{dl}= \frac{1}{2} \Delta_+ + 2\bar{y}\Delta_K +
  \frac{wu}{\sqrt{2}\left(1+u^2\right)},
\end{equation}
consistent with expectations from weak coupling. This implies as
stated earlier, that the critical point will not be exactly at
$\Delta_+=0$.

Of the three leading irrelevant operators, it should be noted that,
although $\bar{y}$ has the same scaling dimension as $\bar{q}$ and
$w$, it has a different role close to the 2CK fixed point. The reason
is that it couples to the term $i\gamma\beta_I$, already present at
the fixed point. But the $\Delta_K$ term at the fixed point suppresses
fluctuations of $\beta_I$, causing the leading term in the
correlations of $\beta_I$ to {\em vanish} at the fixed point (with
sub-dominant terms caused by $g_\gamma\neq 0$). As a result, unlike
$\bar{q}$ and $w$ which each yield $O\left(T\ln T\right)$
contributions to the impurity specific heat\cite{Georges} --- a key
feature of a 2CK non-Fermi liquid --- the singular part arising from
$\bar{y}$ is only of $O\left(T^3\ln T\right)$, for temperatures $T\ll
T_K$. Thus only {\em two} of the above independent leading irrelevant
operators give leading singular specific heat corrections. In fact,
there is a third one, involving only spin degrees of freedom, which is
discussed in Appendix C.

In Section VI, the conditions for accessibility of the 2CK fixed point
are analyzed. We turn here to further analysis of the behavior near
the 2CK fixed point.

\subsection{Symmetry Breaking Operators}
Up to this point we have only dealt with an electron-impurity system
which is invariant under $Z_2$ ($1\leftrightarrow 2$ interchange) and
spin $SU(2)$ symmetry. However, in a realistic situation of an
impurity in a metal, there will generally be a non-zero, although
possibly small asymmetry between the two sites which will break the
$Z_2$ symmetry. Furthermore, in the presence of a magnetic field, the
equivalence between the two spin channels will be lost, leading to a
situation similar to the anisotropic Kondo
problem\cite{Nozieres1,Nozieres2,Coleman}. As might be expected, in
both cases, the symmetry breaking terms are relevant and in their
presence the system flows away from the 2CK fixed point. In this
section, we will briefly comment on the effects of symmetry breaking
terms close to the 2CK fixed point.

It is clear from the discussion in the previous section that for an
operator to be relevant close to the 2CK fixed point, it has to be of
the form $i\delta \chi$ where $\chi$ is a Majorana fermion of scaling
dimension 1/2. From the ten Majorana fermions (four pairs of
$\alpha_\mu$, $\beta_\mu$ and $\gamma$, $\delta$, all listed in Table
\ref{Table1}) we can make nine such operators. Excluding $i\delta
\alpha_{ec}$ and $i\delta \beta_{ec}$ due to total electron number
conservation (which only allows $\alpha_{ec}$ and $\beta_{ec}$ to
appear together as $\alpha_{ec}\beta_{ec}$) we are left with seven
possible terms.

 From the transformation properties under the discrete symmetries of
the system, listed in Table \ref{Table1}, it can be seen that
$\beta_c$ and $\beta_s$ have the same symmetries; indeed this is why
the $\Delta_1$ and $\Delta_-$ terms in Eq(\ref{Hrefermionized1}) could
be combined into the $\Delta_K$ term of Eq(\ref{Hrefermionized2}). In
the presence of small $i\delta\beta_c$ and $i\delta\beta_s$ terms, a
small rotation of the ($\delta$, $\gamma$) pair as well as a small
additional rotation of the ($\beta_c$, $\beta_s$) pair can be
performed to yield just a slightly modified $i\Delta_K\gamma\beta_I$
term, and a single remaining relevant perturbation, the  $i\delta\beta_X$
term. The extra operators ($i\gamma\beta_X$ and $i\delta\beta_I$) are
thus ``redundant''.\cite{Footnote22A} Therefore at the 2CK fixed
point, there are exactly {\em six relevant} operators, all with RG
eigenvalue of 1/2, like $\Delta_+$. These, along with their symmetry
properties, are listed in Table \ref{Table2}.

The first three operators in Table \ref{Table2} do not break the
spin $SU(2)$ symmetry. Among these, the first, our familiar
$i\Delta_+\delta\alpha_c$, is interchange and time-reversal symmetric,
corresponding to the relevant part of the channel pseudo-spin operator
$S_x$. Correspondingly, the second, $i\delta\gamma$, breaks
interchange symmetry but is time reversal invariant, corresponding to a
$S_z$ operator. This will result from simple asymmetry between the
impurity energies of the two sites, i.e. a $\sigma_z$ term in the
original Hamiltonian. Finally, $i\delta\beta_c$ (or, equivalently,
$i\delta\beta_s$) breaks interchange and time reversal which makes it an
imaginary operator, i.e. it is generated by a $S_y$ operator in the
channel sector, corresponding to complex hopping matrix elements.

The relevant spin $SU(2)$ breaking operators are $i\delta\alpha_s$,
$i\delta\alpha_{es}$ and $i\delta\beta_{es}$; these correspond to
joint electron-impurity hops accompanied by a spin flip or carrying
electronic spin. They correspond to combinations of
$\sigma_+\left(c^\dagger_{1\uparrow}c_{2\uparrow} -
c^\dagger_{1\downarrow}c_{2\downarrow} \right)$, $\sigma_+
c^\dagger_{1\downarrow}c_{2\uparrow}$,
$\sigma_+c^\dagger_{1\uparrow}c_{2\downarrow}$ and their hermitian
conjugates and are discussed in Appendix B (see Eq(\ref{Hsf}) and
Eq(\ref{MajoranaHsf})). The first corresponds to the ``flavor''
anisotropy term in the conventional two channel Kondo
model\cite{Nozieres1,Nozieres2,Coleman}, and, being interchange and
time reversal symmetric but odd under spin flip, is induced by a
magnetic field in the $z$-direction. Interestingly, the remaining two
relevant spin $SU(2)$ breaking operators are {\em odd} under the $Z_2$
interchange transformation. This means that in order to adjust them to
zero in a non-zero external magnetic field, one would have to tune
also terms that break the interchange symmetry of the problem, making
any additional novel, finite magnetic field, non--Fermi liquid fixed
points (analogous to that found in zero field for $\Delta_+\gg
\Delta_1$,$\Delta_-$), extremely hard to observe in a system that does
not have interchange symmetry.

\section{Additional  intermediate coupling fixed point}
\label{Additionalfixedpoint}
In the previous section we analyzed the behavior of the system close
to the 2CK fixed point, that corresponds to the limit $\Delta_K \gg
\Delta_+$. There, we showed that $\gamma$, ``half'' of the impurity,
acquired non-trivial dynamics, which essentially gave rise to the
non--Fermi liquid behavior of the system. However, as is evident by
examining Eq(\ref{Hrefermionized2}), it should be, in principle,
possible to get the same type of non--Fermi behavior if the inequality
above were reversed ($\Delta_K \ll \Delta_+$).

Indeed, following the same arguments analyzed above, we see that when
$\Delta_K=\bar{q}=\bar{y}=w=u=0$ and $\Delta_+ \neq 0$ we have a
critical point, which has an extra artificial $U(1)$ symmetry, namely
rotations in the $\left(\beta_I, \beta_X\right)$ plane.  At this
secondary fixed point, $\delta$, the other ``half'' of the impurity,
acquires dynamics, rather than $\gamma$. The operators $\bar{q}$,
$\bar{y}$, $i\gamma\beta_I\Psi_{ec}^\dagger\Psi_{ec}$ and
$i\gamma\beta_X\Psi_{ec}^\dagger\Psi_{ec}$ have scaling dimension 3/2
and thus are irrelevant with RG eigenvalue -1/2. But of these, only
the last two will give singular specific heat corrections
($O\left(T\ln T\right)$).

However, there is an important difference from the primary 2CK fixed
point discussed earlier. About this fixed point, there are {\em two}
relevant operators, consistent with the symmetries of the model,
namely $i\gamma \beta_I$ and $i\gamma \beta_X$ (or, equivalently,
$i\gamma \beta_c$ and $i\gamma \beta_s$), both with dimension 1/2.
These correspond simply to $\Delta_1$ and $\Delta_-$ before the change
of variables (Eq(\ref{defbetaix})) leading to
Eq(\ref{Hrefermionized2}); they can be generated from nonzero $q$ and
$y$ . The existence of two relevant operators is due to the fact that
$\beta_I$ and $\beta_X$ transform the {\em same} way under the
discrete symmetries; thus the artificial $U(1)$ symmetry {\em cannot}
be extended into an $O(2)$ group (as was the case for the 2CK fixed
point).  As a result, this new fixed point is harder to find in the
interchange symmetric case than the primary 2CK fixed point, as it
requires the impurity-single-electron hopping term to be small and the
two-electron- plus-impurity and the simple impurity hopping terms to
be almost exactly equal at the Kondo scale.  It should be noted,
however, that the {\em total} number of relevant, dimension 1/2,
operators around this fixed point is again six, including the above
mentioned ones, together with $i\gamma\delta$ and the three spin
$SU(2)$ symmetry breaking operators, $i\gamma\alpha_s$,
$i\gamma\alpha_{es}$ and $i\gamma\beta_{es}$, as discussed in Section
IV.E and listed in Table \ref{Table2}, indicating that the fixed point
symmetry is again that of the conventional 2CK model. This is
supported by the fact that, just like in the case of the 2CK fixed
point, there are four operators with scaling dimension 3/2, albeit
with completely different symmetries.

Finally, some comments are needed on the the nature of this novel
fixed point. We should first stress the {\em absence} of $\Delta_1$,
the impurity-single-electron hopping term, which together with the
$Q_0$ term (see Eq(\ref{calHint})) would form the conventional
Kondo-like interaction term. Hence, the appearance of non-Fermi liquid
behavior does {\em not} originate from the competition between the
spin up and spin down electrons to form a channel-pseudo-spin singlet
ground state with the impurity, but, rather, in the presence of strong
impurity-electron repulsion ($Q_0\approx 1/2$), from the competition
between bare impurity tunnelling and two-electron-plus-impurity
tunnelling. Formally, there is an analogy with the conventional two
channel flavor-anisotropic Kondo model,\cite{Nozieres2} with
$\Delta_\pm$ playing the role of the Kondo couplings of the two
``flavor'' channels, which can be seen in the left hand side of
Eq(\ref{explanatoryforHbosrotated}) if $\Phi_c$ is substituted by
$\Phi_s$. When $\Delta_0\neq\Delta_2$ one flavor channel couples more
strongly to the impurity therefore screening it alone at low energy,
which results in usual Fermi liquid behavior. However, if
$\Delta_0=\Delta_2$ the flavor anisotropy disappears and the system
flows to a non-Fermi liquid fixed point (provided $\Delta_1$ is zero).
As a result, although this fixed point may be in the same universality
class as the conventional 2CK model, the mechanism that brings it
about is completely different physically.

\section{Accessibility of the two channel Kondo fixed point and Conclusions}
In the previous sections we have shown how the physical operators in
the tunnelling impurity problem behave near the two channel Kondo
fixed point. In particular, we observed that a linear combination,
$\Delta_+=\Delta_0+\Delta_2$, of the bare impurity hopping,
$\Delta_0$, and the impurity-plus-two-electron hopping, $\Delta_2$, is
relevant and drives the system away from this special critical point
resulting in conventional Fermi liquid behavior at low temperatures,
as for the usual one-channel Kondo system. If one could somehow tune
$\Delta_+$, (or one other coupling such as the electronic hopping $y$)
then one might be able to tune through the critical point and find the
2CK non-Fermi liquid behavior at low temperatures in the vicinity of a
critical coupling. Unfortunately, such tuning over an adequate range
is probably difficult to achieve. Thus one probably has to rely on the
hope that a natural regime of couplings will lead to flow under
renormalization close to the 2CK fixed point. Vlad\'{a}r and
Zawadowski\cite{Zawadowski1} appear to suggest that this should be the
case. Unfortunately, more complete analysis implies the converse, that
only for fortuitous reasons would the impurity system --- even without
asymmetry between the sites --- exhibit 2CK behavior at low $T$.

In this last section, we use the weak coupling analysis of Section III
and Appendix A combined with the intermediate coupling analysis of
Section IV and Appendix B, to find criteria for approaching close to
the 2CK fixed point.

Since we are interested in systems in which the Kondo temperature is
much less than the bandwidth, the weak hopping behavior will control
the relative strengths of couplings at the Kondo scale, at which the
first of the hopping terms becomes of order the renormalized bandwidth.
Operators which are irrelevant for weak hopping will flow away rapidly
under renormalization, changing by finite amounts the remaining
parameters, $q$, $y$ and the $\left\{\Delta_i\right\}$, $i=0$, 1, 2.
For example, the complicated hopping-scattering term, $w$, which was
discussed in Section IV and plays a role near the intermediate
coupling fixed point, will be of order the hopping terms
$\left\{\Delta_i\right\}$ or smaller initially and flow away rapidly,
modifying, among other terms, $\Delta_+$ as in Eq(\ref{Delta+rgeq}),
in the process. This will be the main role of such a term and we can
incorporate its effects into a modified ``bare'' $\Delta_0$ and
$\Delta_2$.  We thus start at an energy scale substantially below the
bandwidth at which the important parameters for $Q\in\left[0,1\right]$
are just $q$, $y$ and the $\left\{\Delta_i\right\}$ at this scale, the
irrelevant operators having become small.  The relevant eigenvalues
for the hopping about the zero hopping fixed manifold will be
universal. For small $y$ they are given by Eq(\ref{rgivalues}).

If $Q$ is initially small, i.e. $q\leq 1/2$, $\lambda_+$ will be
substantially larger than the other eigenvalues and thus a particular
linear combination of the $\left\{\Delta_i\right\}$ will grow fast.
Unfortunately, as can be seen from Appendix A, this combination
($\tilde{\Delta}_0$) is the wrong one to yield flow near the 2CK fixed
point as it {\em includes} $\Delta_+$. Only if this combination,
$\tilde{\Delta}_0$, is initially very small relative to a power of
another linear combination of the same $\left\{\Delta_i\right\}$, can
behavior near the 2CK critical point be obtained; furthermore the
criteria become more stringent the lower the Kondo temperature, as
shown in Appendix A.

Better prospects occur when $Q\approx1/2$ (i.e. $q$
small). Unfortunately, even if the one-electron plus impurity hopping
term, $\Delta_1$, were initially much bigger than $\Delta_0$ and
$\Delta_2$, the purely electronic hopping term $y$ --- determined
basically by the spatial separation of the two impurity sites\cite{MF}
--- would combine with $\Delta_1$ to generate the {\em wrong}
combination, $\Delta_+$, of $\Delta_0$ and $\Delta_2$, as in
Eq(\ref{Delta+rgeq}). Thus, again, unless $y$ is small the criteria
from Appendix A are very strict, as could be anticipated from the $y$
dependence of $\lambda_\pm$.

The best prospects are thus for $q$ and $y$ both small so that the
eigenvalues are all comparable. But this is just the condition for the
analysis of Section IV and Appendix B via refermionization to be
valid. To get near the 2CK fixed point, $\Delta_+$ must remain small,
thus we can study the RG equations Eq(\ref{intermediatergeqns}) to
leading order in $\Delta_+$ and the linear combinations of $y$ and
$q$, $\bar{y}$ and $\bar{q}$;
\begin{eqnarray}
\label{intermedrgeqnstext2}
  \frac{d\Delta_K}{dl} &=&
 \frac{\Delta_K}{2}\left(1-\frac{\Delta_K^2}{4\pi}\right)
 \\ \nonumber
\frac{d\bar{q}}{dl} &=& -\frac{\Delta_K^2}{8\pi}\bar{q}
\\ \nonumber
\frac{d\bar{y}}{dl} &=& -\frac{\Delta_K^2}{8\pi}\bar{y}
\end{eqnarray}
and, ignoring $w$ from Eq(\ref{Delta+rgeq})
\begin{equation}
  \label{Delta+intermediatergeq}
  \frac{d\Delta_+}{dl} = \frac{1}{2} \Delta_+ + 2\bar{y}\Delta_K.
\end{equation}

At the intermediate coupling 2CK fixed point, the convention we have
chosen yields $\Delta_K^*=\sqrt{4\pi}$, so that $\bar{y}$ and
$\bar{q}$ have the correct eigenvalues there as well as for weak
coupling ($\Delta_K\approx 0$).

Integrating the above equations we
find that the criterion to flow to the 2CK fixed point is, to leading
order in $\Delta_+$ and $\bar{y}$
\begin{equation}
  \label{critsurfacecriterionw/outw}
\Delta_+ -
  \frac{2\bar{y}\Delta_K}{1-\frac{\Delta_K^2}{4\pi}}
  \ln\left|\frac{\Delta_K^2}{4\pi}\right| =0
\end{equation}
The parameters can be evaluated at any scale where
the intermediate coupling RG equations are valid, i.e. for small
$\Delta_+$, $\bar{y}$ and $\bar{q}$. If these are small at the
starting scale, then their starting values can be used. Note that
since $\Delta_K=\sqrt{\Delta_1^2+\left(\Delta_0-\Delta_2\right)^2}$,
unless $\Delta_0$ and $\Delta_2$ are almost exactly equal and
opposite, $\Delta_K$ will be at least as big as $\Delta_+$ initially
so that $\left|\bar{y}\ln\Delta_K\right|$ needs to be small rather than
just $y$, even in the best case of $q=1/2-Q$ small (recall that
$\bar{y}$ is a linear combination of $q$ and $y$ given by
Eq(\ref{defqybar})).

  We thus see that the flows near the 2CK
fixed point, in particular the generation via Eq(\ref{Delta+rgeq}) of
the unique relevant operator, $\Delta_+$ from other operators, yield
stringent conditions for the accessibility of the non-Fermi liquid 2CK
behavior. If there is no symmetry between the two sites, then the
presence of a second relevant operator (see Section IV) makes
prospects even worse.

This work strongly suggests that to observe two-channel-Kondo-like
non-Fermi liquid behavior of an impurity hopping between two sites in a
metal one must either be able to tune some parameters over a
substantial range, or be extremely lucky. This casts doubt on the
interpretation of Ralph {\em et al}\cite{Ralph1,Ralph2} of their narrow
constriction tunnelling data. One possibility, although perhaps
farfetched, is that these might be some kind of defects tunnelling in
environments with higher symmetry, or at least approximate
symmetry. In another paper, we will show how an impurity hopping among
{\em three} sites with triangular symmetry can, without fine tuning,
lead to a two channel Kondo behavior at low temperatures.

\acknowledgments
We would like to thank Andreas Ludwig, Jinwu Ye, Dan
Ralph, Jan von Delft, Igor Smolyarenko and especially Anirvan Sengupta
for useful discussions. This work was supported by the National
Science Foundation via grant DMR 91-06237.
\end{multicols}
\appendix
\section{Details  of  Weak--Intermediate Coupling Crossover}
In this Appendix  we analyze the dependence of the crossover energy
scale, $T_K$, and other quantities
as a function  of initial values of various coupling constants. For
simplicity, we ignore the coupling to the even parts of the electron
fields (and hence terms like $u$ and $w$ in Eq(\ref{Hbosrotated})). Since
 in the weak coupling RG flow equations (Eq(\ref{weakrgeqns})) of
 $\Delta_0$, $\Delta_1$  and
$\Delta_2$, $q$ and $y$ are essentially
constant, these equations can be trivially integrated. The solutions
for the two quantities of interest at the intermediate coupling
region, i.e. $\Delta_+\left(l\right)$ and $\Delta_K\left(l\right)$ are
\begin{equation}
  \label{Delta+(l)}
  \Delta_+\left(l\right)=\Delta_0\left(l\right)
  +\Delta_2\left(l\right)= \tilde{\Delta}_0e^{\lambda_+l}+
  \tilde{\Delta}_2e^{\lambda_-l}
\end{equation}
and
\begin{equation}
  \label{DeltaK(l)}
  \Delta_K\left(l\right)=\sqrt{\Delta_1^2\left(l\right)+
    \Delta_-^2\left(l\right)}=
  \sqrt{\left(\tilde{\Delta}_1e^{\lambda_0l}\right)^2+
\left(\tilde{\Delta}_0e^{\lambda_+l}-
  \tilde{\Delta}_2e^{\lambda_-l}\right)^2},
\end{equation}
with the weak coupling eigenvalues given by Eq(\ref{rgivalues})
\begin{eqnarray*}
 \lambda_\pm&=&\frac{1}{2} -2q^2_0 \pm 2\sqrt{q_0^2+y_0^2} \\ \nonumber
  \lambda_0&=&\frac{1}{2} -2q^2_0
\end{eqnarray*}
Note that the weak coupling eigenvalues for the relevant hopping terms
should be universal.\cite{Kagan} But since we have ignored higher
order terms in
$y$, (as well as renormalization of $q$ from high energy scales which
make it related to a  phase shift rather than a coupling constant) the
expressions for the eigenvalues in terms of the original $y_0$ and
$q_0$ will not be exact.

The couplings that appear in Eqs(\ref{Delta+(l)}) and
(\ref{DeltaK(l)}), $\tilde{\Delta}_0$, $\tilde{\Delta}_1$ and
$\tilde{\Delta}_2$, are linear combinations of the original
$\Delta_0$, $\Delta_1$ and $\Delta_2$:
\begin{eqnarray}
  \label{tildeDelta}
  \tilde{\Delta}_0&=&\frac{1}{2\sqrt{q^2+y^2}}
  \left\{\left(\sqrt{q^2+y^2}+q\right) \Delta_0
    +\left(\sqrt{q^2+y^2}-q\right) \Delta_2 +y\Delta_1\right\} \\
    \nonumber
\tilde{\Delta}_1&=&\frac{1}{\sqrt{q^2+y^2}}
  \left\{ y\left(\Delta_2- \Delta_0\right) +q\Delta_1\right\} \\
    \nonumber
\tilde{\Delta}_2&=&\frac{1}{2\sqrt{q^2+y^2}}
  \left\{\left(\sqrt{q^2+y^2}-q\right) \Delta_0
    +\left(\sqrt{q^2+y^2}+q\right) \Delta_2 -y\Delta_1\right\}.
\end{eqnarray}
The weak coupling flow equations in Eq(\ref{weakrgeqns}) break down at
the crossover scale $T_K=\tau_c^{-1}e^{-l^*}$, which is reached when
the first of $\Delta_+\left(l^*\right)$ and
$\Delta_K\left(l^*\right)$ becomes of order unity.

At this scale, $\bar{q}$ and $\bar{y}$, defined in Eq(\ref{defqybar})
and appearing in Eq(\ref{Hrefermionized2}) are
\begin{eqnarray}
  \label{qybarintermsofTK}
 \bar{q}&=&\sqrt{q^2+y^2}
\frac{\tilde{\Delta}_1}{\left(T_K\tau_c \right)^{\lambda_0}}
\\ \nonumber
 \bar{y}&=&\sqrt{q^2+y^2}
 \left(\frac{\tilde{\Delta}_0}{\left(T_K\tau_c\right)^{\lambda_+}}
   -\frac{\tilde{\Delta}_2}{\left(T_K \tau_c\right)^{\lambda_-}}\right).
\end{eqnarray}

In order to flow close to the 2CK fixed point, it is necessary that
$\Delta_K\left(l^*\right)=O\left(1\right)$  first, with
$\Delta_+\left(l^*\right)\ll 1$. There are two regions of parameter
space, in terms of $\tilde{\Delta}_0$, $\tilde{\Delta}_1$ and
$\tilde{\Delta}_2$, where this is realizable. The first is
\begin{eqnarray}
  \label{condition1}
  \frac{\left|\tilde{\Delta}_0
  \right|^{\lambda_0}}{\left|\tilde{\Delta}_1\right|^{\lambda_+}}&\ll&
1 \\ \nonumber
\frac{\left|\tilde{\Delta}_2
\right|^{\lambda_0}}{\left|\tilde{\Delta}_1\right|^{\lambda_-}}&\ll&1,
\end{eqnarray}
in which case the crossover scale is
\begin{equation}
  \label{TK1}
  T_K\approx\left|\tilde{\Delta}_1\right|^{\frac{1}{\lambda_0}}\tau_c^{-1}.
\end{equation}

The other region corresponds to the  second square under
the square-root sign in Eq(\ref{DeltaK(l)}) being of order unity. In
this case the inequality of Eq(\ref{condition1}) is reversed,
$\tilde{\Delta}_0$ and $\tilde{\Delta}_2$ have to be
opposite in sign, ($\tilde{\Delta}_0\tilde{\Delta}_2 <0$) and
\begin{equation}
  \label{condition2}
  \left|\frac{\left|\tilde{\Delta}_2\right|^{\lambda_+}}{\left|
      \tilde{\Delta}_0\right|^{\lambda_-}}-1\right| \ll1.
\end{equation}
Then $T_K$ is
\begin{equation}
  \label{TK2}
T_K\approx\left|\tilde{\Delta}_0\right|^{\frac{1}{\lambda_+}}
\tau_c^{-1}\approx
 \left|\tilde{\Delta}_2\right|^{\frac{1}{\lambda_-}} \tau_c^{-1}.
\end{equation}

The opposite limit corresponds to $\tilde{\Delta}_+\left(l^*\right)$
becoming of order unity first. This will happen when
Eqs(\ref{condition2})-(\ref{TK2})  hold, but with $\tilde{\Delta}_0$ and
$\tilde{\Delta}_2$ having the same sign
($\tilde{\Delta}_0\tilde{\Delta}_2 >0$).

It can be seen from Eq(\ref{tildeDelta}) that the above two conditions
Eq(\ref{condition1}) and Eq(\ref{condition2}) are complicated
functions of the initial parameters $q$, $y$ and the $\Delta$'s. In
the remainder of this Appendix we will analyze two limiting cases in
terms $q$ and $y$.

The first limit is $\left|q\right|\gg \left|y\right|$. From
Eq(\ref{tildeDelta}) we see that for $q>0$ this corresponds to
$\tilde{\Delta}_i\approx \Delta_i$ for $i=0$, 1 and 2, in which case
the conditions in terms of $\Delta_i$ are read directly from
Eqs(\ref{condition1})-(\ref{condition2}). It should be noted that
small initial $Q_0=\frac{1}{2}-q_0$, i.e. weak coupling of the
electrons to the impurity position, (Eq(\ref{calHint})), lies in this
regime.

In the
  second limiting case, whence
  $\left|q\right|\ll\left|y\right|$,
  \begin{eqnarray}
    \label{qllylimit}
    \tilde{\Delta}_0&\approx&\frac{1}{2}\left(\Delta_0+\Delta_2+
    sgn\left(y\right) \Delta_1\right)\\ \nonumber
    \tilde{\Delta}_2&\approx&\frac{1}{2}\left(\Delta_0+\Delta_2-
    sgn\left(y\right) \Delta_1\right)\\ \nonumber
\left|\tilde{\Delta}_1\right|&\approx&\left|\Delta_2-\Delta_0\right|.
  \end{eqnarray}
In this region, Eq(\ref{condition1}) becomes, for $y>0$, approximately,
\begin{eqnarray}
  \label{condition1qlly}
  \left|\left(\Delta_0+\Delta_2\right)+\Delta_1\right|&\ll&
    \left|\Delta_0-\Delta_2\right|^{1+4y}\\ \nonumber
\left|\left(\Delta_0+\Delta_2\right)-\Delta_1\right|&\ll&
    \left|\Delta_0-\Delta_2\right|^{1-4y},
\end{eqnarray}
while Eq(\ref{condition2}), for $y>0$, corresponds to
\begin{eqnarray}
  \label{condition2qlly}
  \left|\Delta_1\right|&>&\left|\Delta_0+\Delta_2\right|\\ \nonumber
\left|\left(\Delta_0+\Delta_2\right)-\Delta_1\right|^{1-4y}&\approx&
 \left|\left(\Delta_0+\Delta_2\right)+\Delta_1\right|^{1+4y}.
\end{eqnarray}

In conclusion, for both limiting situations, as well as for the
intermediate region, where $\left|q\right|\approx\left|y\right|$,
there is a finite, albeit small, region of initial parameters where
the approach  to the 2CK fixed point at low energies is indeed possible.

Finally, we demonstrate that the weak coupling condition for flowing
to the 2CK fixed point, $\Delta_+\left(l^*\right)$ small, evaluated at
the Kondo temperature, $T_K=\tau_c^{-1}e^{-l^*}$, matches with the one
derived from the intermediate coupling side
(Eq(\ref{critsurfacecriterionw/outw})) in the parameter region where
both are valid, i.e. small initial $\bar{q}$, $\bar{y}$, $\Delta_+$
and $\Delta_K$. To linear order in $q$, $y$ and $\Delta_+$ we have
from Eq(\ref{DeltaK(l)}), $\Delta_K(l)\approx e^{l/2} \Delta_K$ and
hence
\begin{equation}
  \label{TK=DeltaK2}
T_K\approx\Delta_K^2\tau_c^{-1}.
\end{equation}
The condition for criticality becomes from Eq(\ref{Delta+(l)})
\begin{equation}
  \label{Delta+atTK=0}
\frac{\tilde{\Delta}_0}{\left(T_K\tau_c\right)^{\lambda_+}}
   +\frac{\tilde{\Delta}_2}{\left(T_K \tau_c\right)^{\lambda_-}}=0.
\end{equation}
After expanding the factors $T_K^{-\lambda_\pm}$ to leading order in $q$
and $y$ and writing $l^*\approx -\ln\left(\Delta_K^2\right)$
this yields
\begin{equation}
  \label{expandedDelta+atTK=0}
  \left(\frac{\tilde{\Delta}_0}{\left(T_K\tau_c\right)^{\lambda_0}}
   +\frac{\tilde{\Delta}_2}{\left(T_K
\tau_c\right)^{\lambda_0}}\right)-2\sqrt{q^2+y^2}
\left(\frac{\tilde{\Delta}_0}{ \left(T_K\tau_c\right)^{\lambda_0}}
   -\frac{\tilde{\Delta}_2}{\left(T_K
     \tau_c\right)^{\lambda_0}}\right) \ln \Delta_K^2=0.
\end{equation}
Noting that the terms in the first parentheses is just $\Delta_+$
evaluated at $T_K$, to lowest order in $q$ and $y$, while the term
multiplying the logarithm is just $2\bar{y}$ at $T_K$, to lowest order
in $q$ and $y$ (see Eq(\ref{qybarintermsofTK})), we recover the
condition Eq(\ref{critsurfacecriterionw/outw}) derived from the
refermionization calculation, for the initial $\Delta_K$ small.

\section{Detailed Analysis of Intermediate Coupling Point}
In this Appendix the details of the intermediate coupling point
analysis are discussed. The starting point is the
bosonization of the Hamiltonian in the
Eqs(\ref{calHo}-\ref{calHhop}), using Eq(\ref{Psij(x)}) and
Eq(\ref{dPhidx}), but inserting and keeping track of ordering
factors which ensure proper anticommutation relations between
different fields. The original fermions $c_{1\uparrow}$,
$c_{2\uparrow}$, $c_{1\downarrow}$, $c_{2\downarrow}$ are bosonized in
the following way (rather that just the simple form of Eq(\ref{Psij(x)})):
\begin{eqnarray}
  \label{cisbosonization}
  c_{1\uparrow}&=&\frac{1}{\sqrt{2\pi\tau_c}} e^{i\Phi_{1\uparrow}} \\
  \nonumber
c_{2\uparrow}&=&\frac{1}{\sqrt{2\pi\tau_c}} e^{i\Phi_{2\uparrow}}
e^{i\pi\left(N_{1\uparrow}+N_{2\uparrow}\right)} \\ \nonumber
c_{1\downarrow}&=&\frac{1}{\sqrt{2\pi\tau_c}} e^{i\Phi_{1\downarrow}}
e^{i\pi\left(N_{1\uparrow}+N_{2\uparrow}\right)} \\ \nonumber
c_{2\downarrow}&=&\frac{1}{\sqrt{2\pi\tau_c}} e^{i\Phi_{2\downarrow}}
e^{i\pi\left(N_{1\uparrow}+N_{2\uparrow}+N_{1\downarrow}+N_{2\downarrow}
\right)}
\end{eqnarray}
where $N_{1\uparrow}$, $N_{2\uparrow}$ etc. are the number operators
for the associated fields:
\begin{equation}
  \label{N1uparrow}
  N_{1\uparrow}=\int dx
  c_{1\uparrow}^\dagger\left(x\right)c_{1\uparrow}\left(x\right)
  =\frac{1}{2\pi} \int d\epsilon c_{1\uparrow\epsilon}^\dagger
  c_{1\uparrow\epsilon}
\end{equation}
etc. with the total number of electron fields $N_{tot}=N_{1\uparrow} +
N_{2\uparrow} +N_{1\downarrow}+N_{2\downarrow}$ being always
conserved. Note that since the number operators have integer
eigenvalues the exponential factors in Eq(\ref{cisbosonization}) can
only take the values plus and minus one. With these extra ordering
factors the only change in the effective Hamiltonian of
Eq(\ref{defbosonizedH}) is the appearance of
\begin{equation}
\label{expfactor}
\exp\left[i\pi\left(N_{1\uparrow}+N_{2\uparrow}\right)\right]
\end{equation}
multiplying $y$ and $\Delta_1$. Since the combination
$N_{1\uparrow}+N_{2\uparrow}$ is conserved in the absence of spin-flip
processes, the exponential factor can be set to plus or minus unity
and be disregarded. This can be seen more clearly if, using the number
operators $N_c$, $N_s$, $N_{ec}$ and $N_{es}$ defined in an analogous
way as the corresponding $\Phi$-fields in Eq(\ref{Phiosigma}) and
Eq(\ref{Phiec,s}), we realize that
$N_{1\uparrow}+N_{2\uparrow}=N_{ec}+N_{es}$. Since $N_{tot}=2N_{ec}$
which is always conserved, the exponential factor of
Eq(\ref{expfactor}) will be conserved if $N_{es}$ is constant, i.e. if
no spin-flip processes can appear.  [Note that in the absence of spin
flip processes, one can always choose a convention that different spin
species {\em commute } rather than anti-commute.]

After performing the unitary transformation using
$U=\exp\left[-\frac{i}{2}\sigma_z \Phi_c\right]$ we get the
transformed Hamiltonian $\tilde{\cal H}$ of Eq(\ref{Hbosrotated}).
Next, following Emery and Kivelson\cite{EK}, we refermionize the
exponentials of the Bose fields $\Phi_\mu$ with $\mu=c,s,ec,es$ and
also fermionizing $\sigma_-$ keeping track of ordering operators, as
in Eq(\ref{cisbosonization}), to ensure proper anticommutation
relations between different fermion fields \{$\Psi_\mu$\} and $d$:
\begin{eqnarray}
  \label{Psifermionization}
  \Psi_s&=&\frac{1}{\sqrt{2\pi\tau_c}} e^{i\Phi_s} e^{i\pi N_s} \\
  \nonumber
  \Psi_c&=&\frac{1}{\sqrt{2\pi\tau_c}} e^{i\Phi_c}
  e^{i\pi\left(N_{ec}+N_{es}+N_s+N_d\right)} \\ \nonumber
  d&=& \sigma_-
  e^{i\pi\left(N_s+N_{es}+N_{ec}\right)} \\ \nonumber
  \Psi_{es}&=&\frac{1}{\sqrt{2\pi\tau_c}} e^{i\Phi_{es}}
e^{i\pi\left(N_{es}+N_{ec}+N_s\right)} \\ \nonumber
\Psi_{ec}&=&\frac{1}{\sqrt{2\pi\tau_c}} e^{i\Phi_{ec}} e^{i\pi N_s}
\end{eqnarray}
where $N_\mu$ is the number operator for the $\mu$-fermions and
$N_d=d^\dagger d$. The Hamiltonian now becomes
\begin{eqnarray}
  \label{Hhat}
  \hat{\cal H}= {\cal H}_0 &+& \frac{1}{2\sqrt{2\pi\tau_c}} \left\{
  -\Delta_1\left(d+d^\dagger\right)\left(\Psi_s-\Psi_s^\dagger
\right)+
  \Delta_-\left(d+d^\dagger
\right)\left(\Psi_c-\Psi_c^\dagger\right)\right\}
\\ \nonumber
  &+&2u \Psi_{ec}^\dagger\Psi_{ec} +
  \frac{\Delta_+}{2\sqrt{2\pi\tau_c}}
   \left(d-d^\dagger\right)\left(\Psi_c+\Psi_c^\dagger\right)
\\ \nonumber
&+& \pi y\left(2d^\dagger d-1\right)
\left(\Psi_c+\Psi_c^\dagger\right) \left(\Psi_s-\Psi_s^\dagger \right)
\\ \nonumber
&-&2\pi q \Psi^\dagger_c\Psi_c\left(2d^\dagger d-1\right)-
\frac{\sqrt{2\pi\tau_c} w}{2} \left(d-d^\dagger\right)\left(\Psi_c+
\Psi_c^\dagger\right)
\Psi^\dagger_{ec}\Psi_{ec}.
\end{eqnarray}
Note that we have chosen the ordering operators in  the
refermionization, so that the ordering operators
appearing in the bosonized Hamiltonian cancel.

To simplify the Hamiltonian we introduce a set of eight Majorana
fermions of Eq(\ref{defMajoranafermions}), so that the Hamiltonian
takes the form of Eq(\ref{Hrefermionized1}). Their transformation
properties under the various symmetries are listed in Table
\ref{Table1}. Now, as it is clear from this Table, $\beta_s$ and
$\beta_c$ have the same transformation properties under the discrete
symmetries of the system. Therefore, we can make linear combinations
of the two so that only one will couple to the $\gamma$ Majorana
fermion. Defining
\begin{eqnarray}
  \label{defbetaix}
\beta_I&=&\frac{\Delta_-\beta_c - \Delta_1 \beta_s}{\Delta_K}\\ \nonumber
 \beta_X&=&\frac{\Delta_1\beta_c + \Delta_- \beta_s}{\Delta_K}
   \end{eqnarray}
and
\begin{eqnarray}
  \label{defqybar}
  \bar{q}&=&\frac{q \Delta_1 - y\Delta_-}{\Delta_K}\\ \nonumber
 \bar{y}&=&\frac{q\Delta_- + y\Delta_1}{\Delta_K}
\end{eqnarray}
where   $\Delta_K=\sqrt{\Delta_1^2 + \Delta_-^2}$ as in
Eq(\ref{defDeltaK}), we arrive to the Hamiltonian in the form of
Eq(\ref{Hrefermionized2})
\begin{eqnarray*}
  \hat{\cal H}= {\cal H}_0 &+& 2iu\alpha_{ec}\beta_{ec} +
  \frac{i}{\sqrt{2\pi\tau_c}} \left(\Delta_K
   \gamma \beta_I +
  \Delta_+ \delta\alpha_c \right)\\ \nonumber
 &+& 4\pi\bar{q} \gamma\delta\alpha_c\beta_X +4\pi\bar{y}
 \gamma\delta\alpha_c\beta_I +\sqrt{2\pi\tau_c}w
 \delta\alpha_c\alpha_{ec}\beta_{ec},
\end{eqnarray*}
In the RG  the operator $i\gamma\beta_X$ is
{\em redundant}; under renormalization it can be generated but
can always be rotated away.\cite{Footnote22A}

Before deriving the RG flow equations for the various operator
amplitudes in Eq(\ref{Hrefermionized2}) close to the intermediate
coupling fixed point, we should first locate the fixed point. As
discussed in Section IV, formally it should lie at $\Delta_K=\infty$.
This can be seen in the following way. At the fixed point, with
$\Delta_+=\bar{q}=\bar{y}=w=0$ for convenience, the $\Delta_K$ term in
Eq(\ref{Hrefermionized2}) should have the same scaling dimension as
${\cal H}_0$, i.e. unity. Defining the Fourier transform of the
non-interacting ($\Delta_K=0$), imaginary time Green's functions for
$\gamma$ and $\beta_I$ as
\begin{equation}
  \label{G0gamma}
  G_{0\gamma}\left(i\omega\right)=\frac{1}{i\omega
    g_\gamma}=-F.T. \left<T_\tau
  \left[\gamma\left(\tau\right)\gamma\left(0\right) \right] \right>_0
\end{equation}
and
\begin{equation}
  \label{G0betaI}
  G_{0\beta_I}\left(i\omega\right)=-i\frac{sgn\left(\omega
  \right)}{2}=
  -F.T.
  \left<T_\tau
  \left[\beta_I\left(\tau\right)\beta_I\left(0\right) \right]
  \right>_0,
\end{equation}
respectively, where $g_\gamma$ is related related to the normalization
of $\gamma$, as discussed in Section IV, and $F.T.$ stands for Fourier
transform, we see that the corresponding interacting ($\Delta_K\neq0$)
Green's functions are
\begin{equation}
  \label{Ggamma}
  G_\gamma\left(i\omega\right)=\frac{1}{G^{-1}_{0\gamma}
  \left(i\omega\right) - \frac{\Delta_K^2}{2\pi\tau_c}
  G_{0\beta_I}\left(i\omega\right)} =
\frac{1}{ig_\gamma \omega+ \frac{i\Delta_K^2
    sgn\left(\omega\right)}{4\pi\tau_c} }
\end{equation}
and
\begin{equation}
  \label{GbetaI}
  G_{\beta_I}\left(i\omega\right)=\frac{1}{G^{-1}_{0\beta_I}
  \left(i\omega\right) - \frac{\Delta_K^2}{2\pi\tau_c}
  G_{0\gamma}\left(i\omega\right)} =\frac{1}{2i sgn\left(\omega\right)+
  \frac{i\Delta_K^2}{2\pi\tau_c g_\gamma\omega}}
\end{equation}
This means that the naive scaling dimensions of $\gamma$ and $\beta_I$
($\omega\rightarrow 0$) with $\Delta_K$ and $g_\gamma$ {\em finite}
are 1/2 and 3/2, respectively, adding to 2. Thus some combination of
$\Delta_K$ and $g_\gamma$ must be infinite at the fixed point. But
more useful than letting $\Delta_K\rightarrow\infty$ is to keep
$\Delta_K$ finite and let $g_\gamma\rightarrow 0$.  A convenient form,
which we will adopt, is to choose the normalization of $\gamma$ so
that
\begin{equation}
  \label{DeltaKgconstraint}
  g_\gamma +\frac{\Delta_K^2}{4\pi}=1
\end{equation}
This fixes the scaling dimension of $\gamma$ to be 1/2 at the fixed
point, but also correctly gives the conventional scaling as
$\Delta_K\rightarrow 0$. This choice determines the part of RG flow
equations for $\Delta_K$ and
$g_\gamma$ that does not involve any other parameters:
\begin{eqnarray}
  \label{ggammaDeltaKfloweqns}
  \left.\frac{d\Delta_K}{dl}\right|_{rescaling}&=&\frac{g_\gamma}{2}
  \Delta_K=\frac{\Delta_K}{2}-\frac{\Delta_K^3}{8\pi}\\ \nonumber
\left.\frac{dg_\gamma}{dl}\right|_{rescaling}&=&- \frac{\Delta_K^2}{4
    \pi}
  g_\gamma
\end{eqnarray}
Thus, close to the weak coupling fixed point ($\Delta_K\approx0$,
$g_\gamma$=1), $g_\gamma$ is marginal and $\Delta_K$ is relevant with
dimension 1/2, while at the intermediate coupling fixed point
($\Delta_K$ finite and $g_\gamma\approx 0$), $\Delta_K$ is marginal
and $g_\gamma$ irrelevant.  In addition, the flow equation for
$\Delta_K$ in Eq(\ref{ggammaDeltaKfloweqns}) determines the crossover
scale, i.e. $\Delta_K\left(l^*\right)=O\left(1\right)$, where the
renormalized values of the various amplitudes from the weak and
intermediate coupling parts of the flows can be patched together.

Next, we will derive the RG flow equations close to the 2CK fixed
point. This involves four usual steps. We will use a simple frequency
shell RG, taking the unperturbed Hamiltonian to be $H_0$ with non-zero
$\Delta_K$ and $u$; note that as usual, certain coefficients will
depend on the RG scheme. First the modes with magnitude of their
frequency in the infinitesimal range between $\tau_c^{-1}
\left(1-dl\right)$ and $\tau_c^{-1}$ are integrated out. Then the
  frequency cutoff is rescaled to its initial value $\tau_c^{-1}$ and
  the fields are rescaled by their scaling dimension, determined by
  their two point correlation functions, from which the RG eigenvalues
  of the various operators can be determined. Third, perturbative
  corrections to the flow equations are included from integrating out
  the high energy modes in the four fermion terms, for small
  $\bar{q}$, $\bar{y}$ and $w$. Thus, e.g. the contraction (or
  integration) of $\gamma$ and $\beta_I$ in the $\bar{y}$-term
  contributes to the generation of $\Delta_+$. Finally, since at each
  step of this decimation process $i\gamma \beta_X$ will be generated,
  we have to rotate $\beta_I$ and $\beta_X$ so that it disappears.
  However, the corrections due to this rotation are negligible to the
  order at which we work. Following this procedure we get the full set
  of RG equations describing the 2CK fixed point, and more generally
  for small $\Delta_+$, $w$, $\bar{y}$ and $\bar{q}$, but with general
  $\Delta_K$,
\begin{eqnarray}
  \label{intermediatergeqns}
  \frac{d\Delta_+}{dl} &=& \frac{1}{2} \Delta_+ + 2\bar{y}\Delta_K +
  \frac{wu}{\sqrt{2}\left(1+u^2\right)}
  +O\left(\bar{y}^2\Delta_+,w^2\Delta_+, wu\Delta_+,\Delta_+^3\right)
  \\ \nonumber
 \frac{d\Delta_K}{dl} &=&
 \frac{\Delta_K}{2}\left(1-\frac{\Delta_K^2}{4\pi}\right) +
 2\bar{y}\Delta_+
 +O\left(\bar{y}^2\Delta_K, \Delta_+^3\right)
 \\ \nonumber
\frac{d\bar{q}}{dl} &=& -\frac{\Delta_K^2}{8\pi}\bar{q} +
O\left(\bar{q}^2 \frac{\Delta_+}{\Delta_K},\bar{y}^2 ,
\frac{\Delta_+}{\Delta_K}
\bar{q} \bar{y} \frac{\Delta_+}{\Delta_K}\right)
\\ \nonumber
\frac{d\bar{y}}{dl} &=& -\frac{\Delta_K^2}{8\pi}\bar{y} + O
\left(\bar{q}^2
\frac{\Delta_+}{\Delta_K},\bar{y}^2 \frac{\Delta_+}{\Delta_K}, \bar{q}
\bar{y} \frac{\Delta_+}{\Delta_K}\right)
\\ \nonumber
\frac{dw}{dl} &=& -\frac{w}{2} + O\left(u\Delta_+\right)
\\ \nonumber
\frac{du}{dl}&=&-\frac{w\Delta_+}{4\pi}
\end{eqnarray}
The constraint of Eq(\ref{DeltaKgconstraint}) has been used; it has
the effect of simplifying various messy terms that would appear in
denominators from Eqs(\ref{Ggamma}) and (\ref{GbetaI}) evaluated at
$\left|\omega\right|=1/\tau_c$.\cite{Footnoteonrganisotropy} The above
flow equations are identical to the weak coupling flow equations
Eq(\ref{weakrgeqns}), to lowest order in $y$, $q$, $\Delta_i$ and
$q\Delta_i$ and $y\Delta_i$, with $i=+$, $-$, 1 or $K$. The higher
order terms in the $\Delta_i$'s are non-universal and depend on the RG
procedure. In the flow equations above, we have selected a RG scheme
via Eq(\ref{DeltaKgconstraint}), such that the scaling dimensions of
the various operators are given correctly to leading order in the both
weak and intermediate coupling limits. The $u$ dependence of the $w$
term in $\frac{d\Delta_+}{dl}$ can, however, depend on the RG scheme,
or, equivalently, the definition of $u$.

If $w$, $\Delta_+$ and $\bar{y}$ are all small at some scale, then the
$w\Delta_+$ and $\bar{y}\Delta_+$ terms can be ignored. The RG
equations can then be integrated to yield the condition for lying on
the critical manifold:
\begin{equation}
  \label{critsurfacecriterion}
  \Delta_+ + \frac{wu}{\sqrt{2}\left(1+u^2\right)} -
  \frac{2\bar{y}\Delta_K}{1-\frac{\Delta_K^2}{4\pi}}
  \ln\left|\frac{\Delta_K^2}{4\pi}\right|
    = 0
\end{equation}
to linear order in $w$, $\bar{y}$ and $\Delta_+$. In general, the
parameters above should be evaluated in the regime of the crossover to
intermediate coupling. [In the approximation of ignoring the neglected
terms, the condition Eq(\ref{critsurfacecriterion}) can be used at any
scale.] The role of $w$ in the above equation is a typical example of
the way an irrelevant operator can renormalize the value of a relevant
one before decaying away. In this context, one should remember that
the values of both $\Delta_K$ and $\Delta_+$ should include
corrections from the more strongly irrelevant operators that exist at
higher energy scales but become negligibly small as we approach the
fixed point, or as the flow moves away from weak coupling.

The effects of various symmetry breaking terms are discussed in
Section IV.D. The symmetries of the important operators are summarized
in Table \ref{Table2}. If spin-flipping is allowed, it was shown in
Section IV.D that the relevant processes involve joint
electron-impurity hops which carry electronic spin. Their
representation in terms of the bare fields involves ordering
operators:
\begin{eqnarray}
  \label{Hsf}
  {\cal H}_{sf}= e^{i\pi\left(N_{ec}+N_{es}\right)} \left\{\Delta_{es}
  \left(\sigma_+e^{i\Phi_c}-\sigma_-e^{-i\Phi_c}\right)
    \left(e^{i\Phi_{es}}-e^{-i\Phi_{es}}\right)\right. \\ \nonumber
  -\Delta'_{es} i \left( \sigma_+e^{i\Phi_c}-\sigma_-e^{-i\Phi_c}\right)
    \left(e^{i\Phi_{es}}+e^{-i\Phi_{es}}\right) \\ \nonumber
 +\left.\Delta_s  \left(\sigma_+e^{i\Phi_c}-\sigma_-e^{-i\Phi_c}\right)
    \left(e^{i\Phi_s}-e^{-i\Phi_s}\right) \right\};
\end{eqnarray}
but after refermionizing, these disappear with the choices made in
Eq(\ref{Psifermionization}), yielding, in the Majorana representation,
\begin{equation}
  \label{MajoranaHsf}
  {\cal H}_{sf}= 2\sqrt{2\pi\tau_c}i \left(\Delta_{es} \delta \alpha_{es}
  +\Delta'_{es} \delta \beta_{es} +\Delta_s \delta \alpha_s\right).
\end{equation}

\section{Leading Irrelevant Operators -- Comparison to Conformal Field
  Theory Results} In this paper we have implemented an RG approach
based on a bosonization method first developed by Emery and
Kivelson\cite{EK} which is inherently anisotropic in the channel
pseudo-spin. We thus found a 2CK-like fixed point with an $O(2)$
symmetry, higher than the $Z_2$ interchange symmetry of the original
impurity problem. Despite the fact that this fixed point had lower
symmetry than the $SU(2)$ symmetric one of the 2CK model solved by
conformal field theory methods, we will show in this Appendix that
{\em all} relevant, marginal and leading irrelevant operators expected
from conformal field theory can be recovered from this
model.\cite{Affleck3}

To be specific, in Section IV.D the six dimension 1/2 relevant
operators of the 2CK fixed point were analyzed (see Table
\ref{Table2}). Also, the only marginal, dimension 1, operator was
found to be the conventional potential scattering term,
$u\Psi^\dagger_{ec}\Psi_{ec}$, yielding $i\alpha_{ec}\beta_{ec}$.

In addition, there are four leading irrelevant, dimension 3/2,
operators consistent with the symmetries of the problem we considered.
The first, $\gamma\delta\alpha_c\beta_X$, corresponding to $\bar{q}$,
does not break the $O(2)$ symmetry of the fixed point. This is
essentially the leading irrelevant operator discussed in reference
\onlinecite{Georges}. In fact, there is no other operator with the
same dimension,\cite{Senguptathanks} invariant under $O(2)$ and the
other discrete symmetries of the system, that does not involve even
operators ($\Phi_{ec}$, $\Phi_{es}$), which naively decouple from the
impurity. This can be checked directly from Table \ref{Table1}, by
taking all possible combinations of Majorana fermions in the channel
pseudo-spin sector. Since the 2CK fixed point can be correctly
described in terms of the free Majorana fermions we conclude that
there is no other operator missing from our approach. This fact
suggests that the operators that would distinguish the $O(2)$ from the
$SU(2)$ symmetric fixed points have dimension higher than 3/2 and are
thus more strongly irrelevant, in contrast with what Affleck {\em et
  al}\cite{Affleck2} have claimed that a term breaking the $SU(2)$
down to an $O(2)$ symmetry has dimension 3/2.\cite{Gogolin}

The second leading irrelevant operator,
$\delta\alpha_s\alpha_{es}\beta_{es}$, can be seen from Table
\ref{Table1} to be the only other operator with dimension 3/2 which
respects the $O(2)$ as well as the discrete, symmetries of the system.
Although it explicitly involves the Majorana fermions of the spin
sector, it can be seen that this operator also satisfies the full spin
$SU(2)$ symmetry by the fact that it is generated from a strictly spin
$SU(2)$ invariant term, $\Delta_1$. Starting from Eq(\ref{Hhop1}), one
can first bosonize the fermion operators $c^\dagger_{i\sigma}$
directly into exponentials of combinations of $\Phi_{ec}$,
$\Phi_{es}$, $\Phi_c$ and $\Phi_s$. Then the $\Delta_1$ operator
becomes
\begin{eqnarray}
  \label{Deltabeforefusion}
  \frac{\Delta_1}{4\pi\tau_c}\left\{\sigma_+\left[e^{-\frac{i}{2}
    \left(\Phi_{ec} + \Phi_{es} - \Phi_c -\Phi_s\right)}
    e^{\frac{i}{2}\left(\Phi_{ec}+  \Phi_{es} + \Phi_c
      +\Phi_s\right)}\right.\right.  \\ \nonumber
\left.\left.+ e^{-\frac{i}{2}\left(\Phi_{ec}- \Phi_{es} - \Phi_c
    +\Phi_s\right)}
e^{\frac{i}{2}\left(\Phi_{ec}- \Phi_{es} + \Phi_c -\Phi_s\right)}
\right] + h.c.\right\}
\end{eqnarray}
(multiplied by $\exp i\pi\left(N_{es}+N_{ec}\right)$, the appropriate
ordering operator discussed in Appendix B). Obviously, the leading
operator of this term is the one appearing in Eq(\ref{defbosonizedH}).
However there are subleading corrections, coming from the operator
product expansion of the various exponentials. One such term is
\begin{equation}
  \label{fusionresult}
  i\left(\sigma_+e^{i\Phi_c}- \sigma_- e^{-i\Phi_c}\right)
  \frac{\partial \Phi_{es}}{\partial x} \sin\Phi_s,
\end{equation}
with a cutoff dependent coefficient proportional to $\Delta_1$. This
is allowed by the full symmetries of the 2CK problem and will
therefore be generated in the way outlined above or, equivalently,
from the high energy scattering processes. At weak coupling, this
term, with RG eigenvalue $-1/2-2q^2$, is always irrelevant.  After
performing the unitary transformation of Eq(\ref{Utrans1/2}) and
refermionizing, it becomes $\delta\alpha_s\alpha_{es}\beta_{es}$, with
scaling dimension 3/2 close to the intermediate coupling fixed point.
This operator contributes to the singular part of the impurity
specific heat but not to the susceptibility of the conventional two
channel, spin 1/2, Kondo model, thus leading to a {\em non-universal}
Wilson ratio (The existence of two irrelevant operators at the 2CK
fixed point was discussed by Affleck and Ludwig in Appendix D of
reference \onlinecite{Affleck3}). Other approaches to the 2CK
problem\cite{Georges,EK,Betheansatz} assumed implicitly some higher
symmetry\cite{Senguptaprivcom} {\em at} the fixed point and therefore
do not contain this second leading irrelevant operator (see reference
\onlinecite{Affleck2} for more details).

The remaining two dimension 3/2 operators, $\gamma\delta
\alpha_c\beta_I$ and $\delta\alpha_c \alpha_{ec}\beta_{ec}$, do {\em
  not} respect the $O(2)$ symmetry, but exist on the critical manifold
with the $Z_2$ interchange symmetry, i.e. having set the relevant
operator, corresponding to a linear combination of $\Delta_+$, $w$ and
$\bar{y}$, to zero --- see Eq(\ref{critsurfacecriterion}). They were
discussed in Section IV.C and correspond to couplings $\bar{y}$ and
$w$.  Interestingly, the first does not enter in the leading impurity
specific heat correction of $O(T\ln T)$, since it couples to a
dimension 2 irrelevant operator, $g_\gamma$, leading only to a $T^3\ln
T$ correction to the impurity specific heat.

More generally, in the case where $Z_2$ interchange symmetry is also
broken, there are three more dimension 3/2 irrelevant operators,
namely $\gamma\delta\alpha_{ec} \beta_{ec}$, $i\delta\frac{\partial
  \beta_s}{\partial x}$ and $i\delta\frac{\partial \beta_c}{\partial
  x}$. However, in this case {\em two} relevant operators must be
tuned to be on the critical manifold.

Finally, for completeness, we summarize the leading irrelevant
operators for the additional non-Fermi liquid fixed point discussed in
Section V. Apart from $\bar{q}$ and $\bar{y}$, the operators
$\gamma\beta_I \alpha_{ec}\beta_{ec}$ and $\gamma\beta_X \alpha_{ec}
\beta_{ec}$ also have scaling dimension 3/2. Again, only the last two
contribute to $O(T\ln T)$ to the impurity specific heat.  As in the
above case, in the absence of $Z_2$ symmetry, three additional
dimension 3/2 operators will be present, namely $i\gamma\frac{\partial
  \alpha_c}{\partial x}$, $\gamma\delta \alpha_{ec}\beta_{ec}$ and
$\gamma\alpha_c \beta_c\beta_s$.

%\end{multicols}
%\pagebreak
\begin{table}[htbp]
  \begin{center}
    \leavevmode
    \begin{tabular}{|c||c|c|c|c|}
Fermion &Site Interchange ($Z_2$)  & Time Reversal   &Spin Reversal
&Spin Conservation mod 2
\\ \hline\hline
$\gamma$&
      +&+&+&+ \\ \hline
$\delta$ &
      --&--&+&+ \\ \hline
$\alpha_c$ &
      --&+&+&+ \\ \hline
$\beta_c$&
      +&--&+&+ \\ \hline
$\alpha_s$&
      --&+&--&+ \\ \hline
$\beta_s$&
      +&--&+&+ \\ \hline
$\alpha_{ec}$&
      +&+&+&-- \\ \hline
$\beta_{ec}$&
      +&--&+&-- \\ \hline
$\alpha_{es}$&
      +&+&--&-- \\ \hline
$\beta_{es}$&
      +&--&+&--
     \end{tabular}
   \end{center}
  \caption{Transformation properties of Majorana fermions with respect
    to the discrete symmetries of the system. Electron number
    conservation implies that the Bose field $\Phi_{ec}$ can be
    shifted by an arbitrary amount, which means that only the
    combination $\alpha_{ec}\beta_{ec}$ can appear. In the presence of
    conservation of the $z$-component of the spin, $\Phi_{es}$ can
    also be shifted by an arbitrary constant. If $z$-spin were only
    conserved mod 2, then just a discrete symmetry would remain; this
    is listed in the last column of the table as a simple way to
    distinguish the different symmetries of the $e$ operators.}
  \label{Table1}
\end{table}

\begin{table}[htbp]
  \begin{center}
    \leavevmode
    \begin{tabular}{|c||c|c|c|c|}
Relevant Operators &Site Interchange ($Z_2$)   &Time Reversal   &Spin
Reversal &Spin Conservation mod 2
\\ \hline\hline
$i\delta\alpha_c$ &
      +&+&+&+ \\ \hline
$i\delta\gamma$&
      --&+&+&+ \\ \hline
$i\delta\beta_X$&
      --&--&+&+ \\ \hline
$i\delta\alpha_s$&
      +&+&--&+ \\ \hline
$i\delta\alpha_{es}$&
      --&+&--&-- \\ \hline
$i\delta\beta_{es}$&
      --&--&+&--
     \end{tabular}
   \end{center}
  \caption{The six independent relevant operators at 2CK fixed point
    and their transformation properties (for last column, see note in
    Table I).  Note that $i\delta\beta_s$ and $i\delta\beta_c$ are not
    listed here but rather the combination $i\delta\beta_X$ which has
    the same symmetries as the other two. (See discussion in Section
    IV.D).}
  \label{Table2}
\end{table}

\begin{multicols}{2}

\end{multicols}

\begin{references}
\bibitem{Footnote1}It turns out that the effects of this artificial
  symmetry remain even if it holds modulo 2 (i.e. $N_1-N_2\rightarrow
  N_1-N_2 \pm 2$).
\bibitem{Footnote2}In the case of  intersite separation that is much
  larger than the inverse Fermi wavevector, both $y$ and the momentum
  dependence of scattering away from the Fermi surface can be regarded
  as small\cite{MF}.
\bibitem{Footnote12A} Throughout this paper, we will be careful to
  keep all terms whose presence, or generation under renormalization,
  plays an important role either for the crossover when some of the
  hopping terms are small, or for the behavior near the 2CK fixed
  point. The other terms, which we neglect, should only give rise to
  no more than (generally non-universal) order one pre-factors.
\bibitem{Footnote22A} A similar situation of an apparent extra, but
  redundant relevant operator, appears in conventional $\phi^4$
  theory: the $\phi^3$ operator, in the presence of a magnetic field
  term $h\phi$ can be transformed away by a non-singular (albeit
  non-linear) transformation of the field $\phi$.
\bibitem{Footnoteonrganisotropy}It should be noted, as
pointed out in the main text that, due to the inherent anisotropy of
the RG  scheme employed here,  the fixed point Hamiltonian will {\em
  not} have the  channel $SU(2)$ symmetry of the 2CK problem.
\end{references}
\end{document}